\documentclass[10pt,final,journal,twocolumn]{IEEEtran}

\usepackage{hyperref} 
\usepackage[dvips]{graphicx}
\usepackage{float,makeidx,cite}
\usepackage{amssymb,amsmath,amsthm}

\providecommand{\abs}[1]{\lvert#1\rvert}
\newcommand{\ud}{\,\mathrm{d}}

\newtheorem{Propi1}{Proposition}

\newtheorem{Theoi1}{Theorem}
\newtheorem{Theoi2}[Theoi1]{Theorem}
\newtheorem{Theoi3}[Theoi1]{Theorem}
\newtheorem{Theoi4}[Theoi1]{Theorem}
\newtheorem{Theoi5}[Theoi1]{Theorem}
\newtheorem{Theoi6}[Theoi1]{Theorem}

\title{Adaptive Transmission in Cellular Networks: Fixed-Rate Codes with Power Control vs Physical Layer Rateless Codes}
\author{Amogh Rajanna, \IEEEmembership{Member, IEEE}, and Carl P. Dettmann
\thanks{Amogh Rajanna and Carl P. Dettmann are with the School of Mathematics, University of Bristol, UK. Email \{amogh.rajanna@ieee.org, carl.dettmann@bristol.ac.uk\}. This work was supported by the EPSRC grant number EP/N002458/1 for the project \emph{Spatially Embedded Networks}. Part of this work was presented at ICC’18 \cite{Conf1}.}}
\begin{document}
\maketitle
\begin{abstract}
Adaptive transmission schemes are a crucial aspect of the radio design for future  wireless networks. The paper studies the performance of two classes of adaptive transmission schemes in cellular downlink. One class is based on physical layer rateless codes with constant transmit power and the other uses fixed-rate codes in conjunction with power adaptation. Using a simple stochastic geometry model for the cellular downlink, the focus is to compare the adaptivity of fixed-rate codes with power adaptation to that of physical layer rateless codes only. The performance of both rateless and fixed-rate coded adaptive transmission schemes are compared by evaluating the typical user success probability and rate achievable with the two schemes. Based on both the theoretical analysis and simulation results, it is clearly shown that fixed-rate codes require power control to maintain good performance whereas physical layer rateless codes with constant power can still provide robust performance.
\end{abstract}
\section{Introduction}
\label{sec_intro}
\subsection{Motivation}
\label{introA}
\begin{figure}
\centering 
\includegraphics[scale=0.5, width=0.5\textwidth]{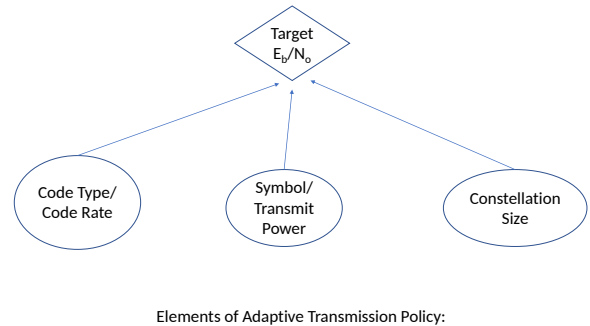}
\caption{Fundamentals of digital transmission of information.}
\label{AMC}
\end{figure}

Adaptive transmission techniques play a key role in the robust design of the radio access network architecture for future cellular networks. The large path loss, high blockage and intermittent (fluctuating) characteristics of the wireless channel at mmWave and higher frequency bands pose a severe bottleneck to the ultra-low latency and ultra-high reliability targeted goals of future networks and the applications they support. The success of future networks in meeting their ambitious goals will certainly be influenced by the performance of adaptive transmission techniques.
The fundamental idea of an adaptive transmission policy in the physical (PHY) layer is to ensure reliable transmission of information bits between the base station (BS) and user in the presence of time-varying channel conditions. This goal can be accomplished by selecting the best possible code type(s), coding rate, constellation size (modulation scheme) and also, by transmit power adaptation to channel conditions as illustrated in Fig.\ref{AMC}. The elements of adaptive transmission policy as in Fig.\ref{AMC} match the rate of transmission adaptively to the instantaneous wireless channel conditions, i.e., the signal to interference noise ratio (SINR). Adaptive transmission has been a well researched topic in the late $1990$'s \cite{Goldsmith,GoldsmithII}, but interesting recent developments in coding theory have led to a renewed focus in this direction\cite{ShokII}.

Rateless codes being a new class of variable-length codes have the innate property to adapt both the code/parity bit construction and also, the number of code/parity bits in response to the time-varying channel conditions. Rateless codes were originally developed for the erasure channel around 15 years ago\cite{Luby}\cite{Shokrollahi}. However, due to the above two properties, rateless codes have been investigated for the noisy channel of the PHY layer too\cite{Bonello}\cite{Soljanin}. From a coding theory perspective, the design and analysis of PHY layer rateless codes over the noisy channel has been a much researched topic recently \cite{Amrit,Amritc,Tian,Kuo} with different design criterion yielding rateless codes with unique and remarkable properties.

In an adaptive transmission policy, the interaction between the code type/rate and transmit power control is further detailed below. In a fixed-rate coded downlink channel, the adverse effects of wireless channel impairments such as small scale fading and path loss can be partially overcome by transmit power control based on channel quality information (CQI) at the BS. This power control leads to an improvement in SINR and thus, coverage probability and rate of transmission. Fixed-rate coding in conjunction with transmit power control based on CQI for maintaining good $E_b/N_o$ over the wireless channel is used in current 4G cellular systems\cite{LTEBook}.
In a PHY layer rateless coded downlink channel, the adaptivity of both the code bit generation and the number of code bits to the channel conditions leads to an enhanced coverage probability and rate of transmission. The key intuition is that with the innate adaptivity of rateless codes, an adaptive transmission policy purely based on rateless codes and no power control can still provide target performance. Such a design holds the potential for enhanced energy-efficiency.

\subsection{Related Work}
\label{introB}
From a communication theory view, the work of \cite{RHI,RHII} studies the performance of rateless codes in the PHY layer of cellular downlink channel assuming fixed transmit power and compares it to that of fixed-rate codes with constant power. The coverage probability and rate enhancements on the downlink channel due to a rateless coded PHY relative to fixed-rate codes are quantified. A stochastic geometry model is used for the cellular downlink, i.e., the BSs are modeled by a homogeneous Poisson point process (PPP). It is shown that the rate and coverage gains due to a rateless PHY are heavily dependent on the interferer activity in the wireless channel and also, the user location relative to the BS. The rate gain of a user close to \emph{only} one BS is different from that of a user \emph{equidistant} from three BSs. Similarly, the rate gain of a user with time-varying interference is different from that of a user with interference invariant to time.

In \cite{Wang}, the performance of both cellular uplink and downlink when employing transmit power adaptation in the form of fractional power control (FPC) is studied. Assuming a stochastic geometry model for the cellular network, the authors consider a scenario where the transmit power is adapted to partially invert the path loss impairment of the wireless channel. The metric used is the coverage probability conditioned on the point process. The conditional coverage probability provides a detailed statistical probe into the performance enhancements for users in cellular uplink and downlink when FPC is employed. In other words, it provides fine-grained information on the distribution of improvements rather than just the average improvement across the network. The key role of transmit power in the analysis and optimization of cellular network performance is detailed in \cite{DiRenzo}. A new analytical framework in which the network throughput depends explicitly on the transmit power of BSs is presented. The key focus is to optimize the energy efficiency (EE) of the network, defined as the ratio of the network throughput to the network power consumption. Refer to \cite{DiRenzo} for a detailed overview of the research efforts dedicated to EE optimization of cellular networks. The results of our current paper will have direct implications on the EE of a cellular network.

\subsection{Contributions}
\label{introC}
The key objective of our paper is to compare the performance of rateless codes with constant power to that of fixed-rate codes employing transmit power adaptation in noisy cellular downlink channels. The wireless channel impairments in the PHY layer of future cellular networks are best captured by using a stochastic geometry model\cite{AndrewsTC,ElSawyII,BHKM2018}. For analytical simplicity and tractability, the common stochastic geometry model for cellular downlink, i.e., a homogeneous PPP for the locations of BSs, is considered in the paper. Adaptive transmission based on
fixed-rate coding uses power control schemes such as channel thresholding, truncated channel inversion and fractional power control to mitigate the channel impairments.
The adaptive transmission scheme based on rateless coding with constant power is compared against multiple fixed-rate coding based schemes. The adaptive schemes are compared under a transmission mode of the cellular downlink where every BS transmits a $K$-bit information packet to its served user. The coverage probability and rate of transmission are the metrics of interest to study the performance.

The technical contributions of our current paper are given below. A clear difference between the current paper contributions and those of prior work \cite{RHI,Wang} is highlighted whenever appropriate.
\begin{enumerate}
  \item We characterize the performance of rateless codes with constant power in cellular downlink via an upper bound on the distribution of the packet transmission time of rateless codes. Note that \cite{RHI} also addresses the performance of rateless codes with constant power in cellular downlink. But in our current paper, the derived analytical results are more accurate, i.e., in terms of the tightness of the bounds and its match to the simulation as illustrated in the numerical results.
  \item The asynchronous mode of operation of cellular downlink is studied through the use of a space time uniform PPP model. The analytical results allow a comparison of the synchronous and asynchronous modes of operation of cellular downlink.
  \item The performance of fixed-rate coded downlink with pathloss-based fractional power control, truncated channel inversion and channel thresholding are quantified.
  \item For the sake of completeness, the paper also provides an analytical characterization of the performance of fading-based channel thresholding and truncated channel inversion in cellular downlink. The metric used is the spatial average coverage probability.
\end{enumerate}
Note that \cite{Wang} also studies the performance of power control in cellular downlink, but focuses on pathloss-based schemes only and on coverage probability conditioned on the point process.

The paper shows that fixed-rate coding with power adaptation maintains good performance only in the low coverage (reliability) regime, i.e., for small values of the delay constraint $N$. On the other hand, rateless coding with constant power performs well in both the low and high reliability (large $N$) regimes. Our results show that fixed-rate coding with power control is inefficient in terms of the cost-benefit tradeoff. While the benefits are coverage and rate gains limited to the low reliability regime, the cost manifests as the system design complexity associated with power adaptation and encoding (decoding) operations. Although power control has played a key role in fixed-rate coded 4G and prior cellular systems, our results demonstrate that PHY layer rateless codes as part of the adaptive transmission policy can provide robust coverage and rate performance without power control.

The organization of the rest of the paper is as follows. Section \ref{sys_mod} contains the system model of the paper and presents the analytical framework used for theoretical study. Section \ref{RatelessCP} characterizes the cellular network performance for rateless codes with constant power. Adaptive transmission schemes based on fixed-rate codes with power adaptation for pathloss and fading are treated in Section \ref{sec_FPC}. Section \ref{asyRless} covers the asynchronous transmission scenario. The numerical results of the paper and the resulting design insights are presented in Section \ref{num_res}. The concluding remarks of the paper are in Section \ref{conc}. The appendices \ref{sec:Theo1}-\ref{sec:Theo6} contain the mathematical derivations.
\section{System Model}
\label{sys_mod}
\subsection{Network Assumptions}
\label{net_mod}
A homogeneous Poisson point process (PPP) $\Phi=\{X_i\}$, $i=0,1,2,\cdots$ of intensity $\lambda$ is used to model the locations of BSs in a cellular downlink setting. We make a simplifying assumption that each BS $X_i$ transmits to one user in its Voronoi cell. The distance between BS $X_i$ and its user located uniformly at random at $Y_i$ is $D_i$. We consider a fixed information transmission mode in which each BS transmits a $K$-bit packet to its user\footnote{We present the sytem model assuming the BSs employ rateless coding. Fixed-rate coding follows as a special case and is presented later in Section IV.A.}. At each BS, a rateless code is used to encode the $K$ information bits. The transmit power of each BS is $\gamma_i$.
The three elements that impair the wireless channel are small scale fading, path loss and interference. The channel has Rayleigh block fading, i.e., the $K$-bit packet is encoded and transmitted within a single coherence time. The packet transmission time of BS $X_i$ to its user $Y_i$ is denoted as $T_i$ symbols (or channel uses). Each $K$-bit packet transmission from a BS is subject to a delay constraint of $N$, i.e., $0< T_i\le N$. For a coherence time $T_c$ secs and signal bandwidth $W_c$ Hz, the value of $N$ is given as $N=T_cW_c$. At time $t\geq 0$, the medium access control (MAC) state of BS $X_i$  is given by $e_i(t)=1\left(0<t\le T_i\right)$, where $1(\cdot)$ is the indicator function.

The received signal at user $Y_i$ at time $t$ is given by
\begin{align}
y_i(t)&=h_{ii}D_i^{-\alpha/2}\sqrt{\gamma_i} x_i(t)+\sum_{k\neq i}g_{ki} \abs{X_k-Y_i}^{-\alpha/2} e_k(t)\sqrt{\gamma_k}\nonumber\\
&~~x_k(t)+z_i,~0<t\le T_i,
\end{align}
where $h_{ii}$ and $g_{ki}$ are the fading coefficients, $\alpha>2$ is the path loss exponent, $x_{i}$ and $x_{k}$ are the transmitted symbols of BSs $X_i$ and $X_k, k\neq i$ respectively, $e_k(t)=1\left(0<t\le T_k\right)$ is the MAC state of BS $X_k, k\neq i$ and $z_i\sim \mathcal{N}_c(0,1)$ is the thermal noise at user $Y_i$. The $1^{\rm st}$ term is the desired signal from BS $X_i$ and the $2^{\rm nd}$ term is the interference from BSs $\{X_k\},k\neq i$. To facilitate an analytical study of the performance of an adaptive transmission policy based on rateless coding, we consider two types of interference models in the cellular downlink. The two models are described below.

\subsubsection{Time-varying Interference}
In this model, the interference power at a user $Y_i$ is function of time $t$.
When the BS $X_i$ is transmitting to its user $Y_i$, all other BSs interfere until they have completed their own $K$-bit packet transmission to their users, i.e., an interfering BS $X_k$ will transmit for a duration of $T_k$ channel uses from $t=0$ and will subsequently turn off. For this case, the instantaneous interference power and SINR at user $Y_i$ at time $t$ are given by\footnote{In our paper, we assume that all BSs have data to transmit at $t=0$. This is a simplifying assumption. Currently, we do not focus on the case where interference is affected by temporal traffic generation models\cite{Gharbieh}.}
\begin{equation}
I_i\left(t\right)=\sum_{k\neq i}\gamma_k\abs{g_{ki}}^2|X_k-Y_i|^{-\alpha}
e_k(t)\label{ct_int}
\end{equation}
and
\begin{equation}
\mathrm{SINR}_i\left(t\right)=\frac{\gamma_i\abs{h_{ii}}^2D_i^{-\alpha}
}{1+I_i(t)},\label{sinr_in}	
\end{equation}
respectively. In (\ref{sinr_in}), the noise power is normalized to 1.

The time-averaged interference at user $Y_i$ up to time $t$ is given by
\begin{equation}
\hat{I}_i(t)=\frac{1}{t}\int_{0}^t I_i(\tau)\ud \tau. \label{Av_Int}
\end{equation}
Since every interferer transmits a $K$-bit packet to its user for $T_k$ symbols and turns off, the interference is monotonic w.r.t $t$, i.e., both $I_i(t)$ and $\hat I_i(t)$ are decreasing functions of $t$.

User $Y_i$ employs a nearest-neighbor decoder based on channel knowledge at receiver only and performs minimum Euclidean distance decoding\cite{Lapidoth}. The achievable rate $C_i(t)$ is given by \footnote{The receiver represented by (\ref{NNdec_tvi}) is a low-complexity practically relevant receiver. Its achievable rate is a lower bound to that of an ideal matched filter receiver. Please see \cite{RHI} for more details.}
\begin{equation}
C_i(t)=\log_2\left(1+\frac{\gamma_i\abs{h_{ii}}^2D_i^{-\alpha}
}{1+\hat{I}_i(t)}\right).\label{NNdec_tvi}
\end{equation}

\subsubsection{Constant Interference}
In this model, we make a simplifying assumption that every interfering BS transmits to their user \emph{continuously} without turning off. The MAC state of an interfering BS $X_k$ at time $t$ is thus given by $e_k(t)=1,~t\geq 0$. Hence, the interference power at the user $Y_i$ does \emph{not} change with time and is given by
\begin{equation}
I_i=\sum_{k\neq i}\gamma_k\abs{g_{ki}}^2|X_k-Y_i|^{-\alpha}\label{co_int}
\end{equation}

The achievable rate $C_i$ in this model is given by
\begin{equation}
C_i=\log_2\left(1+\frac{\gamma_i\abs{h_{ii}}^2D_i^{-\alpha}
}{1+I_i}\right).\label{NNdec_ci}
\end{equation}

The remainder of the discussion presented in this section applies to both the above interference models. Based on (\ref{NNdec_tvi}) and (\ref{NNdec_ci}), the time to decode $K$ information bits and thus, the packet transmission time $T_i$ are given by
\begin{align}
&\hat{T}_i=\min\left\{t:K<t~C_i(t)\right\}\label{GRx_pkt}\\
&T_i=\min(N,\hat{T}_i).\label{pkt_ti}
\end{align}

The distribution of the packet transmission time $T_i$ in (\ref{pkt_ti}) is necessary to characterize the performance of an adaptive transmission policy using rateless codes in a cellular downlink.
\subsection{Typical User Analysis}
\label{theo_ana}
To study the distribution of the packet transmission time, consider the typical user located at the origin. To characterize the complementary CDF (CCDF) of the packet transmission time $T$, we first note that the CCDFs of $T$ and $\hat{T}$ are related as
\begin{align}
\mathbb{P}\left(T>t\right)&=\begin{cases}
\mathbb{P}(\hat{T}>t) & t<N \\
0 & t\ge N.
\end{cases}\label{ccdf_rel}
\end{align}
Next consider the below two events for the constant interference case
\begin{align}
\mathcal{E}_1(t):&~\hat{T}>t\nonumber\\
\mathcal{E}_2(t):&~\frac{K}{t}\geq \log_2\left(1+\frac{\gamma\abs{h}^2D^{-\alpha}
}{1+I}\right),\label{e2_def}
\end{align}
where in (\ref{e2_def}), $I$ is the constant interference at origin given by
\begin{equation}\label{I_eqtn}
I=\sum_{k\neq 0}\gamma_k\abs{g_{k}}^2|X_k|^{-\alpha}.
\end{equation}

For a given $t$, a key observation is that the event $\mathcal{E}_1(t)$ is true if and only if $\mathcal{E}_2(t)$ holds true. This follows from (\ref{GRx_pkt}). Thus
\begin{align}
\mathbb{P}(\hat{T}>t)&=\mathbb{P}\left(\frac{K}{t}\geq \log_2\left(1+\frac{\gamma\abs{h}^2D^{-\alpha}}{1+I} \right)\right)\label{Bap_wri}\\
&=\mathbb{P}\left(\frac{\gamma\abs{h}^2D^{-\alpha}}{1+I}\leq \theta_t \right),
\label{ci_ccdf}
\end{align}
where $\theta_t=2^{K/t}-1$. Assuming a high enough BS density $\lambda$, we ignore the noise term for the remainder of the paper.

Under the time-varying interference case, the CCDF of $\hat{T}$ is given by
\begin{equation}\label{tvi_ccdf}
\mathbb{P}(\hat{T}>t)=\mathbb{P}\left(\frac{\gamma\abs{h}^2D^{-\alpha}}
{\hat{I}(t)}\leq \theta_t \right),
\end{equation}
where $\hat{I}(t)$ is the average interference up to time $t$ at the typical user and is obtained from (\ref{Av_Int}):
\begin{align}
&\hat{I}(t)=\sum_{k\neq 0}\gamma_k\abs{g_{k}}^2|X_k|^{-\alpha}
\eta_{k}(t)\label{avIn_asn}\\
&\eta_{k}(t)=\frac{1}{t}\int_{0}^t 1(\tau\leq T_k)\ud \tau=\min\left(1,T_k/t\right).\label{et_def}
\end{align}
The marks $\eta_{k}(t)$ are correlated for different $k$.

For the $K$-bit packet transmission to the typical user, the performance of the adaptive transmission policy is quantified through the success probability and rate of transmission. The success probability and rate of the typical user are defined as
\begin{align}
&p_{\rm s}(N)\triangleq 1-\mathbb{P}(\hat{T}>N) \label{PsR_exp}\\
&R_{N}\triangleq \frac{Kp_{\rm s}(N)}{\mathbb{E}\left[T\right]}=\frac{Kp_{\rm s}(N)}{\int_{0}^{N}\mathbb{P}(\hat{T}>t)\ud t}. \label{Re_RC}
\end{align}
Note that as per (\ref{pkt_ti}), $T$ is a truncated version of $\hat{T}$ at $N$.
\subsection{Adaptive Transmission Schemes}
\label{AdapTxSch}
In this subsection, we explain the motive for comparing the types of adaptive transmission strategies considered in this paper. For forward error correction (FEC), we consider two scenarios. In one scenario, the cellular network employs rateless codes for FEC while in the second scenario, conventional fixed-rate codes are used for FEC. In an adaptive transmission policy, the code type/rate, symbol power and modulation size can be made adaptive to channel conditions to ensure reliable transmission of bits. In this paper though, we limit the adaptation to only code type/rate and symbol power. Rateless codes have robust adaptivity to channel variations whereas fixed-rate codes do not have the same adaptivity to the channel (see \cite{RHI} for more discussion). Hence for a fair choice of adaptive transmission schemes, we combine rateless codes with constant power and fixed-rate codes with transmit power adaptation. In the following, we discuss the two classes of adaptive transmission policies and quantify their performance.
\section{Rateless Coding with Constant Power}
\label{RatelessCP}
When the cellular network uses rateless codes for FEC, each BS encodes a $K$-bit packet using a variable length code, e.g., a Raptor code or a LT-concatenated code\cite{Bonello} (LT is Luby Transform) with degree distributions optimized for the noisy channel and also, being adaptive to the channel variations. The parity symbols are incrementally generated and transmitted until the $K$ bits are decoded at the user or the maximum number of parity symbols $N$ is reached. The user performs multiple decoding attempts to decode the $K$-bit packet using the Belief Propogation algorithm. An outage occurs if the $K$ bits are not decoded within $N$ parity symbols.

In the first adaptive transmission scheme we consider, the rateless codes are used for FEC and the transmit power is constant, i.e., no power adaptation. Based on (\ref{PsR_exp}) and (\ref{Re_RC}), the success probability and rate of transmission for the $K$-bit packet can be obtained from the CCDF of the typical user packet transmission time. From \cite{RHI}, the CCDF of the packet transmission time under the constant interference model for cellular downlink is given by
\begin{equation}
\mathbb{P}(T>t)= \underbrace{1-\frac{1}
{{}_2F_{1}\left(\left[1,-\delta\right];1-\delta;-\theta_t\right)}}_{P_{\rm c}(t)},~t<N \label{ccdf_CI}
\end{equation}
where $\delta=2/\alpha$ and ${}_2F_{1}\left([a, b]; c; z\right)$ is the Gauss hypergeometric function. Define $\theta=2^{K/N}-1$. The success probability $p_{\rm s}(N)$ can be written as
\begin{equation}\label{Rleps}
p_{\rm s}(N)=\frac{1}
{{}_2F_{1}\left(\left[1,-\delta\right];1-\delta;-\theta\right)}.
\end{equation}
The rate $R_N$ can be obtained based on (\ref{Re_RC}) and (\ref{ccdf_CI}) as
\begin{equation}\label{Rlern}
R_{N}=\frac{K p_{\rm s}(N)}{\int_{0}^{N}P_{\rm c}(t)\ud t}.
\end{equation}

Under the time-varying interference model, the CCDF of the packet transmission time given in (\ref{tvi_ccdf}) does not admit an explicit expression due to the correlated marks in $\hat{I}(t)$ in (\ref{avIn_asn}). In \cite{RHI}, an independent thinning model approximation is proposed to study the dependence of the typical user's transmission time on the time-varying interference of the cellular network.
In the independent thinning model, the correlated marks $T_k$ of (\ref{avIn_asn}) are replaced by i.i.d. $\bar{T}_k$ with CDF $F(\bar t)$. Under this model, the average interference at the typical user is given by
\begin{align}
\bar{I}\left(t\right)&=\sum_{k\neq 0}\gamma_k\abs{g_{k}}^2|X_k|^{-\alpha}
\bar{\eta}_k(t)\label{IA_avin}\\
\bar{\eta}_k(t)&=\min\left(1,\bar{T}_k/t\right).\nonumber
\end{align}
From now onwards, we just use $\bar{\eta}$ instead of $\bar{\eta}(t)$ for brevity.
The typical user packet transmission time $T$ for this model is defined as
\begin{align}
&\hat{T}=\min\left\{t:K<t~\bar{C}(t)\right\}\nonumber\\
&T=\min(N,\hat{T}),\label{pkt_tiITM}
\end{align}
where $\bar{C}(t)$ is the achievable rate based on $\bar{I}(t)$ in (\ref{IA_avin}).

\begin{Theoi1}
\label{Theo1}
An upper bound on the CCDF of typical user packet transmission time under the independent thinning model, $T$ in (\ref{pkt_tiITM}), is given by
\begin{align}
\mathbb{P}\left(T>t\right)&\leq \underbrace{1- \frac{1}{{}_2F_{1} \left(\left[1,-\delta\right]; 1-\delta; -\omega_t\theta_t\right)}}_{P_{\rm s}(t)},~t<N
\label{P_ut}\\
\omega_t&=1-\int_{0}^{1} F(xt) \ud x\label{ome_t}\\
F(t)&=\frac{1}{{}_2F_{1} \left(\left[1,-\delta\right]; 1-\delta; -\theta_t\min\left(1,\mu/t\right)\right)}\nonumber\\
\mu&=\int_0^{N} \left(1-{}_2F_{1}\left(\left[1,\delta\right];1+\delta;-\theta_t\right)\right)\ud t\label{mu_exp}
\end{align}
and $\theta_t=2^{K/t}-1$.
\end{Theoi1}
\begin{IEEEproof}
See Appendix \ref{sec:Theo1}.
\end{IEEEproof}

The success probability for the independent thinning model $\tilde{p}_{\rm s}(N)$ is bounded as\footnote{From (\ref{ome_t}), it can be shown that $\omega(N)=\mathbb{E}[\bar{T}]/N$. A simple substitution for $x$ yields the desired result.}
\begin{equation}\label{tvips}
\tilde{p}_{\rm s}(N)\geq \frac{1}{{}_2F_{1} \left(\left[1,-\delta\right]; 1-\delta; -\theta \mathbb{E}[\bar{T}]/N\right)}.
\end{equation}
The rate $\tilde{R}_N$ can be bounded based on (\ref{Re_RC}) and (\ref{P_ut}) as
\begin{equation}\label{tvirn}
\tilde{R}_N\geq \frac{K\tilde{p}_{\rm s}(N)}{\int_{0}^{N}P_{\rm s}(t)\ud t}.
\end{equation}

When rateless codes are used for FEC, the typical user with constant interference can be interpreted as a user experiencing the worst type of interferer activity in a practical cellular network. Hence, $p_{\rm s}(N)$ and $R_N$ in (\ref{Rleps}) and (\ref{Rlern}) for the constant interference case can be interpreted as a lower bound for the coverage and rate of a practical user in cellular downlink. In a similar way, $\tilde{p}_{\rm s}(N)$ and $\tilde{R}_N$ in (\ref{tvips}) and (\ref{tvirn}) for the time-varying interference case can be interpreted as an approximation for the best (highest) possible rate of a practical user. Rateless coding is able to adapt to changing interference conditions and provide different rates.

\textbf{Remark}: The analytical results obtained in this paper for the case of rateless codes with constant power under the time-varying interference model matches the simulation more accurately than the results in \cite{RHI}.
\section{Fixed-Rate Coding with Power Control}
\label{sec_FPC}
When the cellular network uses fixed rate codes for FEC, each BS encodes a $K$-bit packet using a fixed rate code, e.g., a LDPC code, Turbo code or Reed Solomon code and transmits the entire codeword of $N$ parity symbols. The user receives the $N$ parity symbols over the downlink channel and tries to decode the information packet using the BCJR or Viterbi algorithm. Based on the instantaneous channel conditions, the single decoding attempt can be a success or not.

Pathloss based power control introduces a certain degree of fairness among the users in a cellular downlink. Compensation based on pathloss impairment is more useful in cellular deployments in non-urban settings with large coverage areas. In the following, we quantify the performance when an adaptive transmission scheme employs pathloss based fractional power control (FPC) and fixed-rate coding.
\subsection{Pathloss-based FPC}
\label{FPCplb}
For $0\leq \tau\leq 1$, the transmit power from BS to the typical user is given by \cite{Wang}
\begin{align}
\gamma &=\begin{cases}
\rho/D^{-\tau \alpha}, &~D^{-\alpha}\geq \beta\\
0, &~D^{-\alpha}< \beta.
\end{cases}\label{fpcPL}
\end{align}
The relation between $\beta$ and the max power constraint $\rho_m$ can be obtained based on the fact that $\rho/D^{-\tau \alpha}\leq \rho_m$. The condition $D^{-\alpha}\geq \left(\rho/\rho_m\right)^{\frac{1}{\tau}}$ has to be met to satisfy the maximum power constraint and hence, the threshold $\beta$ is $\beta=\left(\rho/\rho_m\right)^{\frac{1}{\tau}}$.

For fixed-rate coding, the packet transmission time of every BS is fixed to $N$ channel uses and hence, interference is time-invariant as given in (\ref{I_eqtn}). The success probability and rate of the typical user are defined as
\begin{align}
p_{\rm s}(N)&\triangleq \mathbb{P}\left(\mathrm{SIR}>\theta\right) =\mathbb{P}\left(\frac{\gamma\abs{h}^2D^{-\alpha}}{I}>\theta\right)\label{FRC_ps}\\
R_{N}&\triangleq \frac{K}{N}~p_{\rm s}(N)\label{FRC_ra}.
\end{align}

The below theorem quantifies the coverage probability and rate of the typical user under the considered adaptive transmission scheme.
\begin{Theoi2}
\label{The2}
The success probability $p_{\rm s}(N)$ in a cellular downlink employing fixed-rate coding and pathloss-based fractional power control for adaptive transmission is given by
\begin{align}
p_{\rm s}(N)&= \int_{0}^{\frac{\pi \lambda}{\beta^{\delta}}} \exp\left(-z J\left(\theta, z\right)\right) e^{-z} \ud z\label{fpcSP}\\
J\left(\theta, z\right)&=\theta^{\delta} \int_0^{\theta } \frac{\delta}{x^{\delta}} \int_{0}^{\frac{\pi \lambda}{z\beta^{\delta}}} \frac{1}{x+y^{-\tau/\delta}}~z e^{-zy} \ud y \ud x.\label{SPif}
\end{align}
The rate $R_{N}$ can be obtained based on (\ref{FRC_ra}) and (\ref{fpcSP}).
\end{Theoi2}
\begin{IEEEproof}
The proof appears in Appendix \ref{sec:PrfTheo2}.
\end{IEEEproof}
Using the substitution $t=e^{-z}$ simplifies solving the numerical integral in (\ref{fpcSP}) and (\ref{SPif}). The value of $\tau$ can be optimised based on the expression for coverage probability in (\ref{fpcSP}). The value of $\tau$ can be close to $1/2$ for ultra dense networks, uplink and D2D networks, in which interference is a major channel impairment. For networks with less severe interference such as the cellular downlink, the value of $\tau$ can be close to $1$, allowing to benefit from pathloss compensation. The value of the pathloss exponent $\alpha$ influences the FPC exponent $\tau$ chosen.
Rate-splitting decodes and cancels part of the interference and treats the remaining part of the interference as noise, thus bridging the two extremes of fully decoding interference and completely treating interference as noise\cite{Clerckx}.
The use of rate splitting also influences the choice of the value of $\tau$, since rate splitting does partial cancellation of interference.

\subsubsection{Pathloss-based Truncated Channel Inversion}
\label{Tcilb}
The success probability $p_{\rm s}(N)$ in a cellular downlink employing fixed-rate coding and pathloss-based truncated channel inversion for adaptive transmission is given by (\ref{fpcSP}) and (\ref{SPif}) with $\tau=1$. Using numerical results, we illustrate the performance enhancement when $\tau$ changes from $\tau=1$ for full channel inversion to $\tau<1$ for fractional (partial) channel inversion. The truncated channel inversion has the benefit of compensating for the channel impairments with the cost being an increase in the interference power relative to the no channel inversion scheme. In FPC, a balance is achieved between the benefit and cost of channel inversion via a partial compensation of the channel impairments. The partial compensation leads to a decrease in interference relative to the truncated channel inversion, which does full compensation.
\subsection{Pathloss-based Channel Thresholding}
\label{CThrlb}
Now, we focus on a specific form of FPC when $\tau=0$, i.e., the BS transmits with constant power $\rho$ when the user is within a certain range, $D\le \beta^{-1/\alpha}$ as per (\ref{fpcPL}). We quantify the coverage probability and rate of the typical user with pathloss-based channel thresholding.

\begin{Theoi3}
\label{C1fpc}
The success probability $p_{\rm s}(N)$ in a cellular downlink employing fixed-rate coding and pathloss-based channel thresholding for adaptive transmission is given by
\begin{align}
p_{\rm s}(N)&= \frac{1-\exp\left(-\pi \lambda \left(1+H\left(\theta\right) \tilde{F}\left(\beta\right)\right)/\beta^{\delta}\right)} {1+H\left(\theta\right)\tilde{F}\left(\beta\right)}\label{cthPLsp}\\
H(\theta)&=\frac{\theta\delta}{1-\delta}~{}_2F_{1}\left(
\left[1,1-\delta\right];2-\delta;-\theta\right)\label{cthin},
\end{align}
where $\tilde{F}\left(\beta\right)=1-e^{-\pi\lambda/\beta^{\delta}}$.
\end{Theoi3}
\begin{IEEEproof}
The proof is presented in Appendix \ref{sec:PrfTheo3}.
\end{IEEEproof}
The power control schemes based on pathloss compensation have exact closed form characterization and their performance will be discussed in the Section \ref{num_res} on numerical results.
\subsection{Fading-based Channel Thresholding}
\label{sec:RateCThr}
Fading based power control will be useful in small cell and cloud radio access networks with small coverage area. It will also be valuable in indoor wireless networks. The transmit power is adapted based on the value of channel gain $\abs{h}^2$. In channel thresholding, the BS transmits with constant power $\rho$ only if the channel gain $\abs{h}^2$ exceeds a threshold $\beta$ and declares an outage otherwise. Mathematically, the transmit power from BS to the typical user is given by \cite{JindFPC}
\begin{align}
\gamma&=\begin{cases}
\rho, &~\abs{h}^2\geq \beta\\
0, &~\abs{h}^2< \beta.
\end{cases}\label{rho_thres}
\end{align}

The typical user coverage probability and rate are given by (\ref{FRC_ps}) and (\ref{FRC_ra}) together with (\ref{rho_thres}).
\begin{Theoi4}
\label{The4}
The success probability $p_{\rm s}(N)$ in a cellular downlink employing fixed-rate coding and fading-based channel thresholding for adaptive transmission is given by
\begin{align}
p_{\rm s}(N)&\approx \mathcal{F}(\theta)+\mathcal{F}(\theta/\beta) \left[e^{-\beta}-\mathcal{F}(\theta)\right]\label{CThsp}\\
\mathcal{F}(\theta)&=\frac{e^{\beta}}{e^{\beta}-1+
{}_2F_{1}\left(\left[1,-\delta\right];1-\delta;-\theta\right)}.\label{Fexp}
\end{align}
The rate $R_{N}$ can be obtained based on (\ref{FRC_ra}) and (\ref{CThsp}).
\end{Theoi4}
\begin{IEEEproof}
Refer to Appendix \ref{sec:PrfTheo4}.
\end{IEEEproof}
\subsection{Fading-based Truncated Channel Inversion}
\label{sec:FRtci}
In truncated channel inversion (TCI), we adapt the transmit power to invert the channel gain only if the channel gain $\abs{h}^2$ exceeds a threshold $\beta$. Mathematically, the transmit power from BS to the typical user is given by \cite{JindFPC}
\begin{align}
\gamma &=\begin{cases}
\rho/\abs{h}^2, &~\abs{h}^2\geq \beta\\
0, &~\abs{h}^2< \beta.
\end{cases}\label{rho_ci}
\end{align}
The threshold $\beta$ is related to the maximum power constraint $\rho_m$, i.e., $\beta=\rho/\rho_m$.
\begin{Theoi5}
\label{The5}
The success probability $p_{\rm s}(N)$ in a cellular downlink employing fixed-rate coding and fading-based truncated channel inversion for adaptive transmission is given by
\begin{align}
p_{\rm s}(N)&\approx \frac{1}{1+G(\theta)}~e^{-\beta}\label{DiUB}\\
G(\theta)&=\theta^{\delta}\int_0^\theta \frac{\delta}{y^{\delta}}~ e^{y}E_1\left(\beta+y\right)\ud y,\label{Gftn}
\end{align}
where $E_1(x)=\int_{x}^{\infty}e^{-t}/t~\ud t$ is the exponential integral function. The rate $R_{N}$ can be obtained based on (\ref{FRC_ra}) and (\ref{DiUB}).
\end{Theoi5}
\begin{IEEEproof}
The proof appears in Appendix \ref{sec:PrfTheo5}.
\end{IEEEproof}
Note that $G(\theta)$ in (\ref{Gftn}) converges only for $\delta<1$, i.e., $\alpha >2$.
The second term in the RHS of (\ref{DiUB}) represents the loss due to channel truncation while the first term contains the gain due to truncated channel inversion.
The expressions in (\ref{CThsp}) and (\ref{DiUB}) are exact for $\beta=0$ and serve as good approximations for $\beta>0$. The TCI will be useful in downlink scenarios when the channel averaging w.r.t fading cannot be realized. In the paper, we have not discussed the adaptive transmission scheme employing fading-based fractional power control since its theoretical analysis is infeasible.
\section{Asynchronous Transmissions}
\label{asyRless}
In Sections \ref{sys_mod} and \ref{RatelessCP}, the performance of rateless codes in cellular downlink was characterized for the case of synchronous transmissions. In this section, we provide a brief treatment of the performance analysis of rateless codes  under the asynchronous mode of cellular downlink, i.e., BSs transmit packets to their users at different times. The point process model we use to study the asynchronous transmissions case is the so called \emph{Poisson Rain} model, introduced in \cite{Bartek} \cite{nsa-journal}.
A space time homogeneous Poisson point process $\Psi=\{X_k,S_k\},~k=0,1,2,\cdots$ of intensity $\lambda_s$ models the locations of BSs. The parameter $\lambda_s$ is the number of transmission attempts per unit area per unit time. $S_k$ is the packet transmission start time of BS $X_k$. The packet transmission time of BS $X_k$ to its user is $T_k$ symbols. At time $t$, the MAC state of BS $X_k$ is given by $e_k(t)=1\left(S_k\leq t \leq S_k+T_k\right)$.

Similar to the Section \ref{RatelessCP}, we assume an independent thinning model for the analysis of the typical user packet transmission time. Under such a model, each interfering BS $X_k$ transmits for a random duration $\bar{T}_k$ from the start time $S_k$ and becomes silent. The packet times $\bar{T}_k$ are i.i.d. with CDF $F(\bar t)$.
Under this model, the average interference at the typical user is given by
\begin{align}
\bar{I}\left(t\right)&=\sum_{k\neq 0}\gamma_k\abs{g_{k}}^2|X_k|^{-\alpha}
\bar{\eta}_k(t)\label{As_avin}\\
\bar{\eta}_k(t)&=\frac{1}{t}\int_0^t 1(S_k\leq \tau\leq S_k+\bar{T}_k)\ud \tau.\label{et_deas}
\end{align}
The typical user packet transmission time $T$ for this model is defined as $T=\min(N,\hat{T})$ and
\begin{align}
&\hat{T}=\min\left\{t:K<t~\bar{C}(t)\right\},\label{pkt_tias}
\end{align}
where $\bar{C}(t)$ is the achievable rate based on $\bar{I}(t)$ in (\ref{As_avin}).

\begin{Theoi6}
\label{Theo6}
An upper bound on the CCDF of typical user packet transmission time under the independent thinning model, $T$ in (\ref{pkt_tias}), is given by
\begin{align}
\mathbb{P}\left(T>t\right)&\leq \underbrace{1- \frac{1}{{}_2F_{1} \left(\left[1,-\delta\right]; 1-\delta; -\omega_N\theta_t\right)}}_{P_{\rm a}(t)},~t<N
\label{P_aut}\\
\omega_N&=\frac{1}{N}\int_{0}^{N} \left(1-F(t)\right)\ud t\label{ome_N}\\
F(t)&=\frac{1}{{}_2F_{1} \left(\left[1,-\delta\right]; 1-\delta; -\theta_t\min\left(1,\mu/t\right)\right)}\nonumber\\
\mu&=\int_0^{N} \left(1-{}_2F_{1}\left(\left[1,\delta\right]; 1+\delta;-\theta_t\right)\right)\ud t\label{mu_exb}
\end{align}
and $\theta_t=2^{K/t}-1$.
\end{Theoi6}
\begin{IEEEproof}
Refer to Appendix \ref{sec:Theo6}.
\end{IEEEproof}

Below we provide a brief discussion on the performance gap between the synchronous and asynchronous modes of operation in cellular downlink\footnote{The Poisson rain model is an approximate model for the asynchronous mode operation of wireless networks. However, it leads to closed form expressions facilitating comparison with the synchronous case. In the high node density limit, the Poisson rain model approaches the Poisson renewal model, which is a more accurate but complex model for asynchronous transmissions. See \cite{Bartek} \cite{nsa-journal} for more details.}. Note that $P_{\rm s}(t)$ in (\ref{P_ut}) depends on $\omega(t)$ in (\ref{ome_t}). A simple substitution shows that $\omega(t)=\frac{1}{t}\int_{0}^{t}\left(1-F(\nu)\right)\ud \nu$. Since $\omega(t)$ is monotonic, we have $\omega(t)\leq \omega_N$. Hence, both $P_{\rm s}(t)$ and $P_{\rm a}(t)$ from Theorems 1 and 6 satisfy $P_{\rm s}(t)\leq P_{\rm a}(t)$. Based on the definitions in (\ref{PsR_exp}) and (\ref{Re_RC}), it follows that the $p_{\rm s}(N)$ and $R_{N}$ for $P_{\rm s}(t)$, i.e., synchronous mode is lower bounded by that of $P_{\rm a}(t)$, i.e., the asynchronous mode. This observation of performance gain due to synchronization is consistent with \cite{Reed}.
A more thorough discussion on the performance enhancement due to synchronization can be found in \cite{Bartek} \cite{nsa-journal}.
\section{Numerical Results}
\label{num_res}
In this section, we present numerical results that illustrate the performance
benefits of the adaptive transmission policies studied in the paper. The numerical
results provide the performance of the typical user, which is the spatial average
of all users performance in the network. For the simulation, the cellular network
was realized on a square of side $\mathcal{S}=60m$ with wrap around edges. The BS PPP intensity is $\lambda=1m^{-2}$. The information packet size is $K=75$ bits. The cellular network performance was evaluated for varying channel threshold $\beta$
and delay constraint $N$ channel uses/symbols. The acronym CI corresponds to the constant interference model described in Section \ref{sys_mod}. The simulation curve corresponds to the cellular network simulation as per the time-varying interference model described in (\ref{ct_int})-(\ref{pkt_ti}).
The analytical result of \cite[Theorem 2]{RHI} is also plotted and compared against the \emph{tighter} bound result of the current paper, i.e., Theorem 1. In the following figures, the performance of rateless codes with constant power is compared against fixed-rate codes employing power control schemes.
\subsection{Pathloss-based Power Control}
\label{PL_pc}
\subsubsection{Channel Thresholding}
\label{fadcthr}
\begin{figure}[!hbtp]
\centering
\includegraphics[scale=0.5, width=0.5\textwidth]{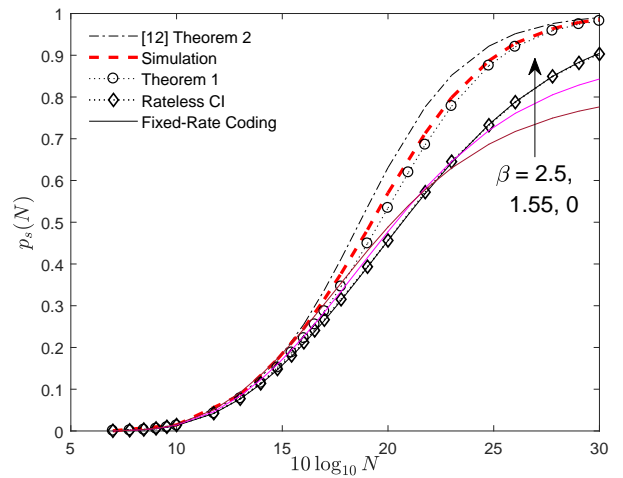}
\caption{Fixed-rate coding with channel thresholding for pathloss: Success probability $p_{\rm s}(N)$ as a function of the delay constraint $N$ in a cellular network with $\lambda=1$ at $\alpha=3$ based on (\ref{Rleps}), (\ref{tvips}), (\ref{FRC_ps}) and (\ref{cthPLsp}) respectively. The solid curves correspond to fixed-rate coding with varying $\beta$. Note that the curves corresponding to rateless CI and fixed-rate $\beta=0$ coincide.}
\label{Psucc_vsN}
\end{figure}
\begin{figure}[!hbtp]
\centering
\includegraphics[scale=0.5, width=0.5\textwidth]{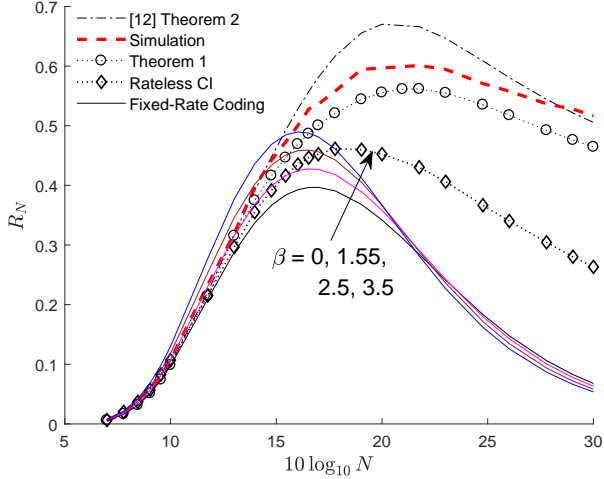}
\caption{Fixed-rate coding with channel thresholding for pathloss: Rate $R_N$ as a function of the delay constraint $N$ in a cellular network with $\lambda=1$ at $\alpha=3$. For fixed-rate coding, the rate is based on (\ref{FRC_ps}), (\ref{FRC_ra}) and (\ref{cthPLsp}). For rateless coding, the rate is obtained from (\ref{Rleps})-(\ref{Rlern}) and (\ref{tvips})-(\ref{tvirn}). The solid curves correspond to fixed-rate coding with varying $\beta$.}
\label{RateNPL}
\end{figure}
Channel thresholding as a power adaptation scheme has both cost and
benefit associated with it. The benefit is that it reduces the interference
for the typical user. The cost being that the serving BS does not transmit
to the user all the time, i.e., only when the pathloss impairment exceeds the
threshold. In Figs. \ref{Psucc_vsN} and \ref{RateNPL}, the success probability $p_{\rm s}(N)$ and rate $R_N$ are plotted as a function of the delay constraint $N$ for both rateless coding with constant power and fixed-rate coding with channel thresholding at $\alpha=3$ and varying threshold $\beta$.
In the high coverage regime, i.e., for large $N$, the cost of not
transmitting to the user under bad channel conditions becomes dominant relative to
the benefit and thus, makes power adaptation inefficient. Hence in this regime,
power adaptation along with fixed-rate coding has no advantages. Rateless
coding with constant power being adaptive to channel conditions, supplies only the necessary number of parity symbols to decode the $K$ bits achieving substantially higher throughput for both the interference models of Section \ref{sys_mod} and hence, the preferred adaptive scheme in this regime.

In the low coverage (or high rate) regime, the benefit of
channel thresholding, i.e., reduced interference allows the BS to transmit
$K$ bits to the user under favorable channel conditions. This benefit offsets
the cost of power adaptation. So fixed-rate coding along with channel
thresholding is useful in the low coverage regime. Rateless coding with no
power adaptation still exhibits good performance due to the fact that expected packet
time (number of parity symbols) is $\mathbb{E}[T]$, unlike non-adaptive fixed-rate coding which always transmits $N$ parity symbols.
\begin{figure}[!hbtp]
\centering
\includegraphics[scale=0.5, width=0.5\textwidth]{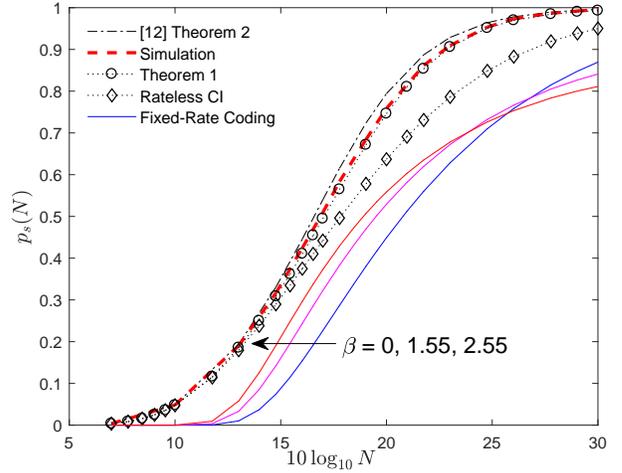}
\caption{Fixed-rate coding with truncated channel inversion for pathloss: Success probability $p_{\rm s}(N)$ as a function of the delay constraint
$N$ in a cellular network with $\lambda=1$ at $\alpha=4$ based on (\ref{FRC_ps}), (\ref{fpcSP}) and (\ref{SPif}) with $\tau=1$. The solid curves correspond to fixed-rate coding with varying $\beta$.}
\label{PsuccNtci}
\end{figure}
\begin{figure}[!hbtp]
\centering
\includegraphics[scale=0.5, width=0.5\textwidth]{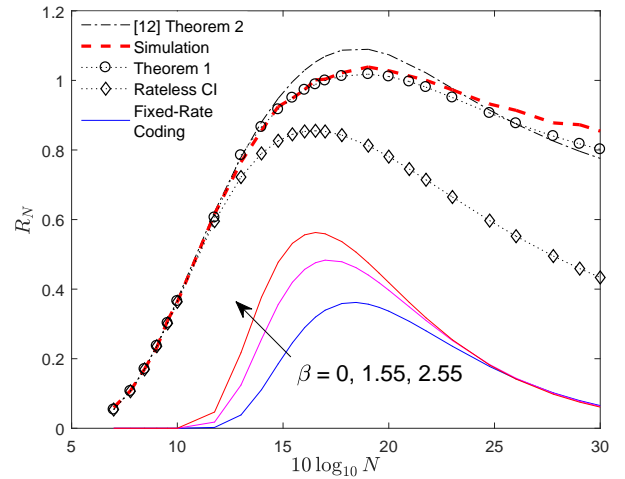}
\caption{Fixed-rate coding with truncated channel inversion for pathloss: Rate $R_N$ as a function of the delay constraint $N$ in a cellular network with $\lambda=1$ at $\alpha=4$. For fixed-rate coding, the rate is based on (\ref{FRC_ps}), (\ref{FRC_ra}), (\ref{fpcSP}) and (\ref{SPif}) with $\tau=1$.}
\label{RateNtciPL}
\end{figure}
\begin{figure}[!hbtp]
\centering
\includegraphics[scale=0.5, width=0.5\textwidth]{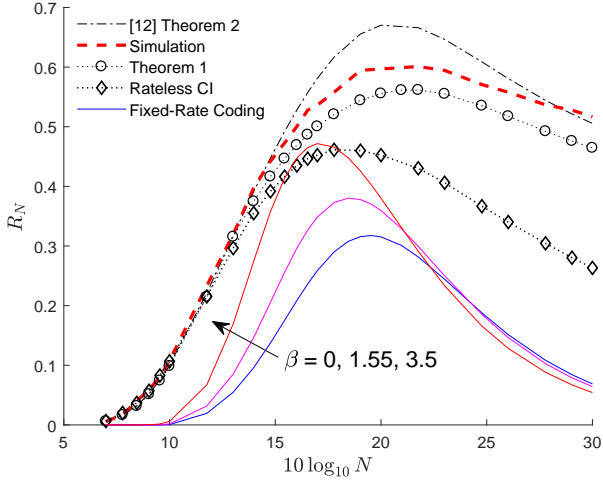}
\caption{Fixed-rate coding with fractional power control for pathloss: Rate $R_N$ as a function of the delay constraint $N$ in a cellular network with $\lambda=1$ at $\alpha=3$. For fixed-rate coding, the rate is obtained from (\ref{FRC_ps}), (\ref{FRC_ra}), (\ref{fpcSP}) and (\ref{SPif}) with $\tau=0.5$. The solid curves correspond to fixed-rate coding with varying $\beta$.}
\label{RateNPLfpc}
\end{figure}
The benefit of channel inversion is that the pathloss from the desired BS to user is compensated for. On the other hand, since the transmit power of interferers is also inversely proportional to the corresponding pathloss, the total interference power at the typical user blows up. The increased interference at the user is the cost of channel inversion. Due to this cost, the thresholding policy $D^{-\alpha}\geq \beta$ is more beneficial in channel inversion compared to the constant power case.
Figs. \ref{PsuccNtci} and \ref{RateNtciPL} show plots of $p_{\rm s}(N)$ and $R_N$
for both rateless coding with constant power and fixed-rate coding with truncated
channel inversion for varying threshold $\beta$. We observe that $\beta=1.55$ provides a substantial increase in both $p_{\rm s}(N)$ and $R_N$ relative to $\beta=0$. (Similar behavior is observed for $\beta=2.55$). For higher values of $\beta$ in Figs. \ref{PsuccNtci} and \ref{RateNtciPL}, we observe the same effect as in the case of channel thresholding for large $N$, i.e., the performance with a higher value of $\beta$ is less than that with a lower value of $\beta$ (around $0$). The curve based on the Theorem $1$ bound for the time-varying interference model has a better match to the simulation curve over a wide range of $N$ relative to the previous bound from \cite[Theorem 2]{RHI}.

\subsubsection{Fractional Power Control}
\label{ptlfpc}
Fig. \ref{RateNPLfpc} shows a plot of the typical user rate $R_N$ for both rateless coding with constant power and fixed-rate coding with pathloss-based fractional power control with $\tau=0.5$ for varying $\beta$.
The channel thresholding scheme transmits at a constant power and does not adversely affect the interference power and the SIR at the typical user. Hence, the success probability and rate with channel thresholding is better than that with either fractional or full truncated channel inversion. When the thresholding $\beta>0$ is applied on a given power control policy, the interferer intensity decreases from $\lambda$ to $\lambda\mathbb{P}(\mathcal{A})$, where $\mathbb{P}(\mathcal{A})=(1-e^{-\pi\lambda/\beta^{\delta}})$. The decrease in interference leads to an increase in SIR and thus, higher success probability and rate at small to moderate $N$. The value of threshold $\beta$ is chosen such that the transmission probability $\mathbb{P}(\mathcal{A})=\{1, 0.9, 0.82, 0.74\}$ which corresponds to $\beta=\{0, 1.55, 2.5, 3.5\}$ for $\lambda=1$.
\subsection{Fading-based Power Control}
\label{fad_pc}
\begin{figure}[!hbtp]
\centering
\includegraphics[width=0.5\textwidth]{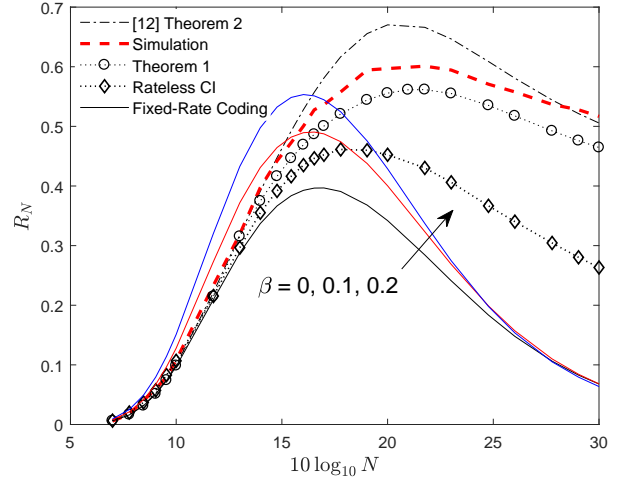}
\caption{Fixed-rate coding with channel thresholding for fading: Rate $R_N$ as a function of $N$ in a cellular network with $\lambda=1$ at $\alpha=3$. For fixed-rate coding, the rate is based on (\ref{FRC_ps}), (\ref{FRC_ra}) and (\ref{CThsp}). For rateless coding, it is based on (\ref{Rlern}) and (\ref{tvirn}). Solid curves represent fixed-rate coding with varying $\beta$.}
\label{Rate_vsN}
\end{figure}

\begin{figure}[!hbtp]
\centering
\includegraphics[scale=0.5, width=0.5\textwidth]{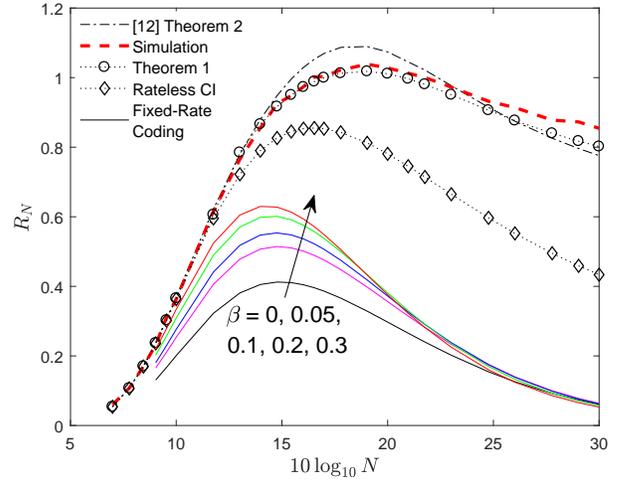}
\caption{Fixed-rate coding with truncated channel inversion for fading: Rate $R_N$ as a function of $N$ in a cellular network with $\lambda=1$ at $\alpha=4$ based on (\ref{FRC_ra}) and (\ref{DiUB}).}
\label{RateNtci}
\end{figure}

Figs. \ref{Rate_vsN} and \ref{RateNtci} show plots of the typical user rate $R_N$ for rateless coding with constant power and fixed-rate coding with fading-based channel thresholding and truncated channel inversion, respectively. The curves for varying $\beta$ are also shown. Figs. \ref{Rate_vsN} and \ref{RateNtci} correspond to the discussions in Sections \ref{sec:RateCThr} and \ref{sec:FRtci}.
The performance gap between rateless coding with constant power and fixed-rate coding with fading-based power control in the above figures is explained by the same discussions as in Section \ref{PL_pc}. When the thresholding $\beta>0$ is applied, the interferer intensity decreases from $\lambda$ to $\lambda\mathbb{P}(\mathcal{A})$, which is $\lambda e^{-\beta}$ for Rayleigh fading. For the cellular network with Rayleigh fading, $\beta=\{0, 0.1, 0.2, 0.3\}$ yields the desired values of $\mathbb{P}(\mathcal{A})$ mentioned previously.

To compare the performance of the power control scheme based on fading to the one based on pathloss, we observe the rate plots of Figs. \ref{Rate_vsN} and \ref{RateNPL} (or Figs. \ref{RateNtci} and \ref{RateNtciPL}). For both sets of plots, we observe that the fading-based power control is more effective, i.e., it has a better performance over a broader range of $N$. We note that for a BS density $\lambda=1$, Rayleigh fading is a much severe channel impairment relative to pathloss. In a cellular network with power control based on pathloss only, the receiver has to cope with the adverse effects of both interference and fading. In the case of power control based on fading only, the receiver has to cope with interference and pathloss, which is milder relative to fading at BS density $\lambda=1$. Hence, fading-based power control has a better rate over a broad range of $N$ relative to pathloss-based power control.
\subsection{System Design Implications}
\label{sdi}
From Figs. \ref{PsuccNtci} and \ref{RateNtciPL}, we observe that for $N=100$ rateless coding achieves a $R_N$ from $0.78$ to $1$ and for $N=300$, a $R_N$ performance of $0.6$ to $0.9$ is achieved. On the other hand, for fixed-rate coding with channel inversion (or fractional power control), a very good $R_N$ can be obtained at $N=100$ by choosing $\beta\geq 2.5$. For power control, the value of $\beta$ needs to be optimized for $N$. At $N=300$ even with optimal $\beta$, the performance gap between rateless coding and fixed-rate coding is too large. Thus, to achieve a desired performance of $p_{\rm s}(N)$ and $R_N$, a fixed-rate coded system has to use channel thresholding along with an optimal $\beta^*(N)$ and this incurs a significant system complexity relative to rateless coding with constant power. For a $K$-bit packet transmission, rateless coding with no power control can achieve good $E_b/N_o$ for the $K$ bits with a higher probability and also, a higher rate relative to fixed-rate coding with power adaptation. Since fixed-rate codes are not adaptive, the transmit power needs to be adapted to maintain an acceptable $E_b/N_o$ for the adaptive transmission of $K$ bits. On the other hand, the robust adaptivity of rateless codes enables good performance even without power adaptation.
\section{Conclusion}
\label{conc}
In this paper, we study two classes of adaptive transmission schemes with the goal of achieving a good $E_b/N_o$ for reliability over the wireless channel of future networks. We compare the performance of rateless coding with constant power to that of fixed-rate coding with power adaptation such as channel thresholding, truncated channel inversion and fractional power control. It is shown that rateless coding with constant power performs much better relative to fixed-rate codes with power control in the moderate to high coverage (reliability) regime. Only in the low coverage regime, the performance of the latter can be made comparable or better than rateless codes by optimal choice of the channel threshold $\beta$. However, these improvements in the low coverage regime come with the cost of power adaptation and feedback resources. Note that the adaptive modulation and coding (AMC) scheme adaptively chooses the code-rate of a fixed-rate code based on the instantaneous channel conditions. As an extension of the current line of research, \cite{RDII} characterizes the performance of AMC with fixed power relative to physical layer rateless codes. Studying the performance of rateless codes with power control can also be a subject of future research.

\section*{Acknowledgement}
The authors like to thank the editor Dr. Marios Kountoris and the reviewers for their constructive comments. The paper material and presentation was greatly enhanced by their careful review. Thanks are also due to Dr. Michael Luby of ICSI, Berkeley, CA for valuable discussions on physical layer rateless coding in wireless communications.
\appendices
\section{Proof of Theorem 1}
\label{sec:Theo1}
First, we recall few expressions from \cite[Appendix C]{RHI} which are used here to derive new results. Note that Appendix C in \cite{RHI} also aims to derive an upper bound on the CCDF of the typical user packet transmission time. But in this appendix, we provide more accurate and tighter bounds relative to \cite[Appendix C]{RHI}.

From \cite[Eqtn (66)]{RHI}, the CCDF of $\hat T$ is given by
\begin{align}
\mathbb{P}\left(\hat{T}>t\right)&= \mathbb{E}\left[
1-\exp\left(-\pi\lambda H(t)D^2 \right)\right]\nonumber\\ &=1-1/\left(H(t)+1\right).\label{ccf_ITM}
\end{align}%

From \cite[Eqtn (65)]{RHI}, we have the following definition
\begin{align}
H(t)&\triangleq \delta \theta_t^{\delta} \mathbb{E}\left[\int_0^{\theta_t} \left(1-\frac{1}{1+\bar{\eta}y}\right) \frac{1}{y^{1+\delta}}\ud y\right].\label{H_ftn}\\
&=\delta \theta_t^{\delta} \mathbb{E}\left[\int_0^{\theta_t} \frac{\bar{\eta}}{\left[1+y\bar{\eta}\right]y^{\delta}} \ud y\right] \nonumber\\
&=\frac{\theta_t\delta}{1-\delta}~\mathbb{E}\left[\bar{\eta} ~{}_2F_{1}
\left(\left[1,1-\delta\right];2-\delta;-\theta_t\bar{\eta}\right)\right]\label{ex_pre}.
\end{align}

The following hypergeometric identity will be useful to simplify $H(t)+1$.
\begin{align}
\frac{\delta}{1-\delta} \theta~&{}_2F_{1}
\left(\left[1,1-\delta\right];2-\delta;-\theta\right)+1\nonumber\\
&\equiv {}_2F_{1} \left(\left[1,-\delta\right]; 1-\delta; -\theta \right).\label{Hyp_id}
\end{align}

Based on (\ref{ex_pre}) and (\ref{Hyp_id}), $H(t)+1$ in (\ref{ccf_ITM}) can be written as
\begin{align}
&\mathbb{P}\left(\hat{T}>t\right)=1-\frac{1}{\mathbb{E}\left[{}_2F_{1} \left(\left[1,-\delta\right]; 1-\delta; -\theta_t \bar{\eta}\right)\right]}\label{ccdf_exp}\\
&~~~\stackrel{(a)}{\leq} 1-\frac{1}{{}_2F_{1} \left(\left[1,-\delta\right]; 1-\delta; -\theta_t \mathbb{E}\left[\bar{\eta}\right]\right)}\label{ccf_Hu}\\
&~~~\stackrel{(b)}{\leq} 1-\frac{1}{{}_2F_{1} \left(\left[1,-\delta\right]; 1-\delta; -\theta_t \min\left(1,\mathbb{E}[\bar{T}]/t\right)\right)}\label{ccf_Hub},
\end{align}
where (a) follows from the concavity of ${}_2F_{1} \left(\left[1,-\delta\right]; 1-\delta;y\right)$ which can be verified easily using
\begin{equation*}
	\frac{\ud}{\ud y} {}_2F_{1}\left(\left[a,b\right];c;y\right)=\frac{a b}{c} {}_2F_{1}\left(\left[a+1,b+1\right];c+1;y\right)
\end{equation*}
and (b) follows from the fact that $\bar \eta=\min\left(1,\bar{T}/t\right)$ is a concave function of $\bar{T}$. In the upper bound of (\ref{ccf_Hu}), the $\mathbb{E}\left[\cdot\right]$ is evaluated as
\begin{align}
\mathbb{E}\left[\bar{\eta}\right]&=\int_{0}^{1}
\mathbb{P}\left(\bar{\eta}>x\right)\ud x \stackrel{(c)}{=}\int_{0}^{1} \mathbb{P}\left(\bar{T}>xt\right)\ud x \label{Eeta},
\end{align}
where in (c), $\bar{\eta}>x$ $\Leftrightarrow$ $\bar{T}/t>x$ since $\bar \eta=\min\left(1,\bar{T}/t\right)$. Based on (\ref{ccf_Hu}) and (\ref{Eeta}), the CCDF of typical user packet transmission time $T$ is given by
\begin{align}
&\mathbb{P}\left(T>t\right)\leq 1- \frac{1}{{}_2F_{1} \left(\left[1,-\delta\right]; 1-\delta; -\omega(t)\theta_t\right)},~t<N\nonumber\\
&\omega(t)=\int_{0}^{1} \mathbb{P}\left(\bar{T}>xt\right) \ud x\label{EetII}.
\end{align}

In (\ref{EetII}), the distribution of interferer packet transmission time $\bar{T}$ is necessary to compute the upper bound in (\ref{ccf_Hu}).
The three possible choices for the distribution of $\bar{T}$ are the upper bound, lower bound and the approximation to the CCDF of the typical user transmission time from \cite{RHI}. Note that as per the Section II.A, the typical cell and interfering cell packet transmission times are identically distributed. Hence, using the upper and lower bounds to the CCDF of typical user transmission time given in \cite{RHI} for the distribution of $\bar{T}$ in (\ref{EetII}) leads to an overestimation and underestimation of interference in (\ref{IA_avin}). For the interference in (\ref{IA_avin}) to accurately model (capture) the interference of the exact model in (\ref{avIn_asn}), we use the CCDF approximation given in \cite[Theorem 2]{RHI} for the distribution of $\bar{T}$ in (\ref{EetII}), i.e.,
\begin{align}
&\mathbb{P}\left(\bar{T}>t\right)=1-\frac{1}{{}_2F_{1} \left(\left[1,-\delta\right]; 1-\delta; -\theta_t\min\left(1,\mu/t\right)\right)}\label{Tbacd}\\
&\mu=\int_0^{N} \left(1-{}_2F_{1}\left(\left[1,\delta\right];1+\delta;-\theta_t\right)\right)\ud t. \label{muxpII}
\end{align}
Plugging the above CCDF of $\bar{T}$ into (\ref{EetII}) with '$xt$' as the sample value of $\bar{T}$ completes the proof.
\section{Proof of Theorem 2}
\label{sec:PrfTheo2}
Define an event $\mathcal{A}:D^{-\alpha}\geq \beta$. Similar to (\ref{sir_c}), the CCDF of $\mathrm{SIR}$ can be written as
\begin{equation}\label{sirfpc1}
\mathbb{P}\left(\mathrm{SIR}>\theta\right)=\underbrace{\mathbb{P}\Big(\frac{\rho\abs{h}^2
D^{-\alpha\left(1-\tau\right)}}{I}>\theta\mid \mathcal{A} \Big)}_{P_1(\theta)}\mathbb{P}(\mathcal{A}).
\end{equation}

Since $D\sim$ Rayleigh $(1/\sqrt{2\pi\lambda})$, we get
\begin{align}
F(d)&=\mathbb{P}\left(D\leq d\right)=1-\exp(-\pi \lambda d^2).\label{fpc1A}
\end{align}
$P_1(\theta)$ in (\ref{sirfpc1}) is evaluated as
\begin{align}
P_1(\theta)
&=\mathbb{P}\Big(\abs{h}^2\geq\frac{\theta I D^{\alpha\left(1-\tau\right)}}
{\rho}\mid \mathcal{A}\Big)\nonumber\\
&\stackrel{(a)}{=}\mathbb{E}\left[\mathcal{L}_I\left(\frac{\theta D^{\alpha\left(1-\tau\right)}}
{\rho}\right)\mid \mathcal{A}\right]\label{p1expA},
\end{align}
where $\mathcal{L}_I(s)=\mathbb{E}\left[e^{-sI}\right]$ is the Laplace transform of interference $I$ by conditioning on $D$ and (a) follows by taking the $\mathbb{E}[\cdot]$ operation w.r.t $I$ by conditioning on $D$. Below we obtain an expression for $\mathcal{L}_I(\cdot)$. Note that in the expression for $I$ in (\ref{I_eqtn}), $\gamma_k$ is the transmit power from BS $X_k$ to its user $Y_k$ and follows the same policy as (\ref{fpcPL}).
\begin{align}
&\mathcal{L}_{I}(s)=\exp\left(-\pi\lambda\mathbb{E}_{\gamma,g}\left[
\int_D^{\infty} \left(1-e^{-s\gamma\abs{g}^2v^{-\alpha}}\right)\ud v^2\right]\right)\nonumber\\
~~~&=\exp\left(-\pi\lambda \int_D^{\infty} \left(1- \mathbb{E}\left[
 e^{-sv^{-\alpha}\gamma\abs{g}^2}\right]\right)\ud v^2\right)\label{LIeq}.
\end{align}
Let $c=sv^{-\alpha}$, the $\mathbb{E}\left[\cdot\right]$ in the integral of (\ref{LIeq}) becomes
\begin{align}
\mathbb{E}\left[e^{-c\gamma\abs{g}^2}\right]&=\mathbb{E}\left[\frac{1}{1+c\gamma}\right]
=\mathbb{E}\left[\frac{1}{1+c\gamma}\mid \mathcal{A}\right]\mathbb{P}(\mathcal{A}) +\mathbb{P}(\mathcal{\bar{A}})\label{Egama}.
\end{align}

The RV $\gamma$ in (\ref{Egama}) represents the transmit power from an interfering BS to its user. To differentiate the transmit power of the BS serving the typical user from that of an interfering BS, we use $D_x$ to denote the distance between an interfering BS and its served user. Recall that $D$ represents the distance between the typical user and its serving BS. Although $D$ and $D_x$ are identically distributed, they are not independent but are assumed so in the following for simplicity due to weak correlation \cite{Wang}. Note that $D$ and $D_x$ follow the Rayleigh distribution in (\ref{fpc1A}). Now (\ref{Egama}) can be written as
\begin{equation}\label{Egama_n}
\mathbb{E}\left[e^{-c\gamma\abs{g}^2}\right]=1-\mathbb{P}(\mathcal{A}) \left(1- \mathbb{E}\left[\frac{1}{1+c\rho D_x^{\tau \alpha}}\big | \mathcal{A}\right]\right).
\end{equation}

Using (\ref{Egama_n}) back in (\ref{LIeq}) and denoting $\kappa=\pi\lambda \mathbb{P}(\mathcal{A})$, we get
\begin{equation}
\mathcal{L}_{I}(s)=\exp\left(-\kappa \int_D^{\infty} \left(1- \mathbb{E}\left[\frac{1}{1+sv^{-\alpha}\rho D_x^{\tau \alpha}}\big | \mathcal{A}\right]\right) \ud v^2\right)\label{Lefpc}
\end{equation}

Plugging $s=\theta D^{\alpha\left(1-\tau\right)}/\rho$ in (\ref{Lefpc}) and using the substitution $y=\theta(D/v)^{\alpha}$, we get
\begin{align}
&\mathcal{L}_{I}\left(\frac{\theta D^{\alpha\left(1-\tau\right)}}
{\rho}\right)=\exp\Big(-\kappa \int_{\theta}^{0} \Big(1- \mathbb{E}\Big[\frac{1}{1+y (D_x/D)^{\tau \alpha}}\nonumber\\
&~~~\big | \mathcal{A}\Big]\Big) \ud y^{-\delta}\theta^{\delta} D^2\Big)\equiv \exp\left(-\kappa M(\theta)\right)\label{Lieq}.
\end{align}
Note that $M(\theta)$ in (\ref{Lieq}) can be written as
\begin{align}
M(\theta)&=\theta^{\delta} D^2\int_0^{\theta} \left(1- \mathbb{E}\left[\frac{1}{1+y (D_x/D)^{\tau \alpha}}\big | \mathcal{A}\right]\right)\frac{\delta}{y^{1+\delta}}\ud y\nonumber\\
&=\theta^{\delta} D^2\int_0^{\theta} \mathbb{E}\left[\frac{1}{y+ (D_x/D)^{-\tau \alpha}}\big | \mathcal{A}\right]\frac{\delta}{y^{\delta}}\ud y.\label{LefpcII}
\end{align}
In (\ref{LefpcII}), the $\mathbb{E}\left[\cdot\right]$ is evaluated similar to
\begin{align}
\mathbb{E}\left[\frac{1}{y+ D_x^{-\tau \alpha}}\big | \mathcal{A}\right]&=
\int_{\mathcal{A}} \frac{1}{y+d_x^{-\tau \alpha}}~\ud F\left(d_x\big | \mathcal{A}\right)\nonumber\\
&=\frac{1}{\mathbb{P}(\mathcal{A})}\int_{0}^{\frac{1}{\beta^{\delta/2}}} \frac{1}{y+d_x^{-\tau \alpha}}\ud F\left(d_x\right)\label{coexDx},
\end{align}
where $F\left(d_x\right)$ is the CDF of distance $D_x$ given in (\ref{fpc1A}). Using (\ref{coexDx}) back in (\ref{LefpcII}) and the resulting $M(\theta)$ in (\ref{Lieq}), we get
\begin{align}
&\mathcal{L}_{I}\left(\frac{\theta D^{\alpha\left(1-\tau\right)}}
{\rho}\right)=\exp\Big(-\pi\lambda D^2 \theta^{\delta} \int_0^{\theta} \frac{\delta}{y^{\delta}} \int_{0}^{\frac{1}{\beta^{\delta/2}}}\nonumber\\ &~~\frac{1}{y+(d_x/D)^{-\tau \alpha}}\ud F\left(d_x\right)\ud y\Big)
\stackrel{(a)}{\equiv} \exp\left(-\pi\lambda D^2 J\left(\theta, D\right)\right)\nonumber,
\end{align}
where (a) basically gives the definition of $J\left(\theta, D\right)$. Hence, we can obtain the CCDF of $\mathrm{SIR}$ as
\begin{align}
&\mathbb{P}\left(\mathrm{SIR}>\theta\right)=\mathbb{E}\left[\exp\left(-\pi\lambda D^2 J\left(\theta, D\right)\right)\big | \mathcal{A}\right]\cdot\mathbb{P}(\mathcal{A})\label{spFpc}\\
&~=\frac{1}{\mathbb{P}(\mathcal{A})}\int_{0}^{\frac{1}{\beta^{\delta/2}}}
\exp\left(-\pi\lambda d^{2} J\left(\theta, d\right)\right) \ud F\left(d\right)\cdot\mathbb{P}(\mathcal{A})\label{sirFpc},
\end{align}
where (\ref{sirFpc}) is obtained using steps similar to (\ref{coexDx}). Hence,
\begin{align}
&p_{\rm s}(N)= \int_{0}^{\frac{1}{\beta^{\delta/2}}} \exp\left(-\pi\lambda d^{2} J\left(\theta, d\right)\right) \ud F\left(d\right),\label{fpcde}\\
&J\left(\theta, d\right)=\theta^{\delta} \int_0^{\theta} \frac{\delta}{x^{\delta}} \int_{0}^{\frac{1}{\beta^{\delta/2}}} \frac{1}{x+(d_x/d)^{-\tau \alpha}}\ud F\left(d_x\right)\ud x\label{deII}.
\end{align}

In (\ref{fpcde}) and (\ref{deII}), we use $z=\pi \lambda d^2$ and $y=(d_x/d)^2$. Noting the changes $\ud F(d)=e^{-z} \ud z$ and $\ud F(d_x)=ze^{-zy} \ud y$, we get the desired form in (\ref{fpcSP}).
\section{Proof of Theorem 3}
\label{sec:PrfTheo3}
The derivation is based on the steps outlined in Appendix \ref{sec:PrfTheo2}. Here, we point out the key differences for $\tau=0$. The CCDF of $\mathrm{SIR}$ is given in (\ref{sirfpc1}). For $\tau=0$, $P_1(\theta)$ can be written as
\begin{equation}\label{cth1a}
P_1(\theta)=\mathbb{E}\left[\mathcal{L}_I\left(\frac{\theta D^{\alpha}}
{\rho}\right)\mid \mathcal{A}\right].
\end{equation}

The equivalent expression of (\ref{Egama_n}) with $\tau=0$ is given by
\begin{equation}\label{Egacth}
\mathbb{E}\left[e^{-c\gamma\abs{g}^2}\right]=1-\mathbb{P}(\mathcal{A}) \left(1- \frac{1}{1+c\rho}\right).
\end{equation}
An expression for $\mathbb{P}(\mathcal{A})$ appears below
\begin{equation}\label{PAeq}
\mathbb{P}(\mathcal{A})=\mathbb{P}(D\leq \beta^{-1/\alpha})=1-e^{-\pi\lambda/\beta^{\delta}}.
\end{equation}
Using $\kappa=\pi\lambda \mathbb{P}(\mathcal{A})$, the $\mathcal{L}_{I}(s)$ is expressed as
\begin{equation}\label{Liscth}
\mathcal{L}_{I}(s)=\exp\left(-\kappa \int_D^{\infty} \left(1-\frac{1}{1+s\rho v^{-\alpha}}\right)\ud v^2\right).
\end{equation}
As in Appendix \ref{sec:PrfTheo2}, plugging the value of $s$ with $\tau=0$ yields
\begin{equation}\label{Lcth}
\mathcal{L}_{I}\left(\frac{\theta D^{\alpha}}{\rho}\right)
=\exp\Big(-\kappa D^2 \underbrace{\theta^{\delta} \int_0^{\theta} \frac{\delta}{\left(1+y\right)y^{\delta}} \ud y}_{H(\theta)}\Big).
\end{equation}

Based on (\ref{Lcth}), the CCDF of $\mathrm{SIR}$ can be written as
\begin{align}
\mathbb{P}\left(\mathrm{SIR}>\theta\right)&=\mathbb{E}\left[\exp\left(-\pi\lambda D^2 \mathbb{P}(\mathcal{A})H(\theta)\right)\mid \mathcal{A}\right]\cdot\mathbb{P}(\mathcal{A})\nonumber\\
&=\int_{\mathcal{A}} \exp\left(-\pi\lambda d^2 \mathbb{P}(\mathcal{A})H(\theta)\right) \ud F(d)\nonumber\\
&\stackrel{(a)}{=}\int_{0}^{\pi\lambda/\beta^{\delta}}e^{-\left(\mathbb{P}(\mathcal{A}) H(\theta)+1\right)y}  \ud y\label{cthApp},
\end{align}
where (a) follows from the substitution $y=\pi\lambda d^2$. Solving the integral in (\ref{cthApp}) and using (\ref{PAeq}) yields the desired result in (\ref{cthPLsp}). An efficient form of $H(\theta)$ defined in (\ref{Lcth}) appears in (\ref{cthin}).
\section{Proof of Theorem 4}
\label{sec:PrfTheo4}
Define an event $\mathcal{A}:\abs{h}^2\geq \beta$. For $\theta>0$, the CCDF of SIR in (\ref{FRC_ps}) is given by
\begin{align}
\mathbb{P}\left(\mathrm{SIR}>\theta\right)&=\mathbb{P}\left(\mathrm{SIR}>\theta, \mathcal{A} \right)+\mathbb{P}\left(\mathrm{SIR}>\theta,\mathcal{\bar{A}} \right)\label{sir_c}\\
&\stackrel{(a)}{=}\underbrace{\mathbb{P}\Big(\frac{\rho\abs{h}^2D^{-\alpha}}
{I}>\theta\mid \mathcal{A} \Big)}_{P_1(\theta)}\mathbb{P}(\mathcal{A})\label{eqsir},
\end{align}
where (a) follows since the $2^{\rm nd}$ term in (\ref{sir_c}) is zero. For $P_1(\theta)$ in (\ref{eqsir}), the conditional CCDF of $\abs{h}^2$ is given by
\begin{align}
\mathbb{P}\left(\abs{h}^2>x\mid \abs{h}^2\geq \beta\right)&=\begin{cases}
e^{-x}/e^{-\beta}, &~x\geq \beta\\
1, &~x< \beta.
\end{cases}\label{codis}
\end{align}
Using (\ref{codis}), the probability $P_1(\theta)$ is expressed as
\begin{align}
P_1(\theta)&=\begin{cases}
\mathbb{E}\left[e^{-\theta D^{\alpha}I/\rho}\right]/e^{-\beta}, &~\theta D^{\alpha}I/\rho \geq \beta\\
1, &~\theta D^{\alpha}I/\rho < \beta.
\end{cases}\label{coprob}
\end{align}
Using $P_1(\theta)$ from (\ref{coprob}) and $\mathbb{P}\left(\mathcal{A}\right)=e^{-\beta}$ in (\ref{eqsir}), we get
\begin{align}
\mathbb{P}\left(\mathrm{SIR}>\theta\right)&=\mathbb{E}\left[e^{-\theta D^{\alpha}I/\rho}\right] \mathbb{P}\left(\theta D^{\alpha}
I/\rho>\beta\right)+\nonumber\\
&~~~\mathbb{P}\left(\theta D^{\alpha}I/\rho<\beta\right) e^{-\beta}\nonumber\\
&=\mathbb{E}\left[e^{-\theta D^{\alpha}I/\rho}\right]+\mathbb{P}\left(\theta D^{\alpha}I/\rho<\beta\right)\nonumber\\
&~~~\left(e^{-\beta}-\mathbb{E}\left[e^{-\theta D^{\alpha}I/\rho}\right]\right)\label{p1_lb}\\
\mathbb{E}\left[e^{-\theta D^{\alpha}I/\rho}\right]&=
\mathbb{E}\left[\mathcal{L}_I\left(\theta D^{\alpha}/\rho
\right)\right]\label{LpT}.
\end{align}
To compute $\mathcal{L}_I\left(\cdot\right)$ in (\ref{LpT}), we note that $\gamma_k$ in the expression for $I$ in (\ref{I_eqtn}) follows the same power policy as in (\ref{rho_thres}). An expression for $\mathcal{L}_I\left(s\right)$ is given in (\ref{LIeq}). To evaluate the $\mathbb{E}\left[\cdot\right]$ in (\ref{LIeq}), let $c=sv^{-\alpha}$. Then
\begin{align}
\mathbb{E}\left[e^{-c\gamma\abs{g}^2}\right]&=\sum_{\mathcal{A},\mathcal{\bar{A}}}
\mathbb{E}\left[e^{-c\gamma\abs{g}^2}\mid i\right] \mathbb{P}\left(i\right)\label{Eexp}\\
&=\mathbb{E}\left[e^{-c\rho\abs{g}^2}\right]\mathbb{P}\left(\mathcal{A}\right)
+\mathbb{P}\left(\mathcal{\bar{A}}\right)\nonumber\\
&\stackrel{(a)}{=}1-e^{-\beta}\left(1-\mathbb{E} \left[e^{-c\rho\abs{g}^2}\right]\right),\label{EXgch}
\end{align}
where (a) uses $\mathbb{P}\left(\mathcal{A}\right)=e^{-\beta}$. Using (\ref{EXgch}) leads to
\begin{equation}
\mathcal{L}_{I}(s)=\exp\left(-\pi\lambda \int_D^{\infty} \left(1- \mathbb{E}\left[ e^{-sv^{-\alpha}\rho\abs{g}^2}\right]\right)\ud v^2 e^{-\beta}\right)\label{expLID}.
\end{equation}
The exponent in (\ref{expLID}), except for $e^{-\beta}$ is identical to the one which results when BSs use constant power\cite{RHI}. The $e^{-\beta}$ factor is due to channel thresholding. Using $\mathcal{L}_{I}(s)$ for the constant power case \cite{RHI} and plugging $s=\theta D^{\alpha}/\rho$, we get
\begin{align}
&\mathcal{L}_{I}(\theta D^{\alpha}/\rho)=\exp\left(-\pi\lambda D^2 H(\theta)e^{-\beta}\right)\label{LIbe}\\
&H(\theta)=\frac{\theta\delta}{1-\delta}~{}_2F_{1}\left(
\left[1,1-\delta\right];2-\delta;-\theta\right).
\end{align}
Taking $\mathbb{E}\left[\cdot\right]$ of (\ref{LIbe}) w.r.t $D\sim$ Rayleigh $(1/\sqrt{2\pi\lambda})$, we get\footnote{In \cite{Hletter,Wang}, a similar distribution has been used for the downlink distance $D$, i.e., $D\sim$ Rayleigh $(1/\sqrt{2\pi c\lambda})$ with $c=1.25$. The factor $c>1$ accounts for the larger area of the Crofton cell relative to the typical cell. However, for simplicity and to be consistent with the previous work \cite{RHI}, we choose  $c=1$. Even though the value of $c$ varies in the literature, it is emphasized that the value of $c$ does not alter the results of the paper.}
\begin{equation}\label{lbcht}
\mathbb{E}\left[\mathcal{L}_I\left(\theta D^{\alpha}/\rho
\right)\right]=\frac{1}{1+H(\theta)e^{-\beta}}\triangleq\mathcal{F}(\theta).
\end{equation}
Based on (\ref{p1_lb}) and (\ref{LpT}), the CCDF of SIR is written as
\begin{equation}\label{ciDiexp}
\mathbb{P}(\mathrm{SIR}>\theta)=\mathcal{F}(\theta)+\mathbb{P}\left(\frac{\theta D^{\alpha}I} {\rho}<\beta\right)\left[e^{-\beta}-\mathcal{F}(\theta)\right].
\end{equation}
\begin{Propi1}
\label{Prp1}
The distribution of $I$ in the RHS of (\ref{ciDiexp}) can be approximated as
\begin{equation}\label{Disap}
\mathbb{P}\left(\frac{\theta D^{\alpha}I} {\rho}<\beta\right)
\approx \mathcal{F}(\theta/\beta).
\end{equation}
\end{Propi1}
\begin{IEEEproof}
The CDF of $I$ in (\ref{Disap}) can be rewritten as
\begin{align}
\mathbb{P}\left(\frac{\theta D^{\alpha}I}{\rho}<\beta\right)
&=\mathbb{P}\left(\frac{\theta D^{\alpha}I}{\beta \rho}<
\mathbb{E}\left[\abs{h}^2\right]\right)\label{Eqpr}.
\end{align}
Consider the two RVs $I$ and $\abs{h}^2$ in (\ref{Eqpr}). The RV $I$ given in (\ref{I_eqtn}) is the dominant RV and mostly determines the scaling of the probability value. On the other hand, $\abs{h}^2$ is the minor component since
$\abs{h}^2\sim$ Exp(1) is a simple RV with $\mathbb{E}\left[\abs{h}^2\right]=1$ and PDF $=e^{-x},~x\in [0,\infty)$. Hence, (\ref{Eqpr}) can be approximated accurately as
\begin{align}
&\mathbb{P}\left(\frac{\theta D^{\alpha}I}{\beta \rho}<
\mathbb{E}\left[\abs{h}^2\right]\right)\approx
\mathbb{P}\left(\frac{\theta D^{\alpha}I}{\beta \rho}<
\abs{h}^2\right)=\nonumber\\
&\mathbb{E}\left[\exp\left(-\frac{\theta D^{\alpha}I}{\beta \rho}\right)\right]=\mathbb{E}\left[\mathcal{L}_I\left(\frac{\theta D^{\alpha}}{\beta \rho}\right)\right]\stackrel{(a)}{=}
\mathcal{F}(\theta/\beta)\label{Sat},
\end{align}
where (a) follows from (\ref{lbcht}).
\end{IEEEproof}
Using (\ref{Sat}) in (\ref{ciDiexp}), the $p_{\rm s}(N)$ can be approximated as in (\ref{CThsp}). The expressions for $\mathcal{F}(\theta)$ in (\ref{lbcht}) and (\ref{Fexp}) are related by the hypergeometric identity in (\ref{Hyp_id}).
\section{Proof of Theorem 5}
\label{sec:PrfTheo5}
Below we obtain the CCDF of the $\mathrm{SIR}$ in (\ref{FRC_ps}) based on the definition of $\gamma$ in (\ref{rho_ci}). Similar to (\ref{eqsir}), the CCDF of $\mathrm{SIR}$ can be written as
\begin{equation}\label{frcti}
\mathbb{P}\left(\mathrm{SIR}>\theta\right)=\mathbb{P}\Big(\frac{\rho D^{-\alpha}}{I}>\theta \Big) \mathbb{P}(\mathcal{A})
\end{equation}
(Note that $\abs{h}^2$ does not appear in the RHS of (\ref{frcti})). Defining $P_1(\theta)$ similar to (\ref{eqsir}), we get
\begin{align}
P_1(\theta)=\mathbb{P}\left(\frac{\rho D^{-\alpha}}
{I}>\theta \right)&=\mathbb{P}\left(\frac{\theta D^{\alpha}I}
{\rho}<1\right)\nonumber\\
&\stackrel{(a)}{\approx}\mathbb{E}\left[\mathcal{L}_I\left(\theta D^{\alpha}
/\rho\right)\right]\label{P1fp},
\end{align}%
where (a) follows by using the same approximation as in Proposition \ref{Prp1} with $\beta=1$. Now (\ref{frcti}) can be written as
\begin{equation}\label{tciDi}
\mathbb{P}\left(\mathrm{SIR}>\theta\right)\approx \mathbb{E}\left[\mathcal{L}_I\Big(\frac{\theta D^{\alpha}}
{\rho}\Big)\right] e^{-\beta}.
\end{equation}

To evaluate $\mathcal{L}_{I}(\cdot)$ in (\ref{P1fp}), we use (\ref{LIeq}) and (\ref{Eexp}). Applying the same steps from (\ref{Eexp})-(\ref{EXgch}) for channel inversion, we get
\begin{align}
\mathbb{E}\left[e^{-c\gamma\abs{g}^2}\right]&=1-e^{-\beta}\left(1-\mathbb{E} \left[e^{-c\rho\abs{g}^2/\abs{h}^2}\mid \mathcal{A}\right]\right)\label{Eetci}\\
&=1-e^{-\beta}\left(1-\mathbb{E} \left[\frac{1}{1+c\rho/\abs{h}^2}\mid \mathcal{A}\right]\right)\nonumber\\
&=1-e^{-\beta}~\mathbb{E}\left[\frac{c\rho}{c\rho+\abs{h}^2}\mid
\mathcal{A}\right].\label{Etc}
\end{align}
Now plugging the value of $c$, (\ref{Etc}) can be rewritten as
\begin{align}
1-\mathbb{E}\left[e^{-sv^{-\alpha}\gamma\abs{g}^2}\right]&=e^{-\beta}~
\mathbb{E}\left[\frac{1}{1+\abs{h}^2/s\rho v^{-\alpha}}\mid \mathcal{A}\right]\nonumber\\
&\stackrel{(a)}{=}\int_{\beta}^{\infty}\frac{1}{1+x/s\rho v^{-\alpha}}e^{-x}\ud x ,\label{CEtc}
\end{align}%
where (a) follows from (\ref{codis}). Using (\ref{CEtc}) back in (\ref{LIeq}) and substituting $s=\theta D^{\alpha}/\rho$, we get
\begin{align}
&\mathcal{L}_{I}\Big(\frac{\theta D^{\alpha}}
{\rho}\Big)=\exp\left(-\pi\lambda \int_D^{\infty}
\int_{\beta}^{\infty}\frac{e^{-x}}{1+xv^{\alpha}/\theta D^{\alpha}}\ud x
\ud v^2\right) \nonumber\\
&\stackrel{(a)}{=}\exp\left(-\pi\lambda \int_{\theta}^{0}
\int_{\beta}^{\infty}\frac{1}{1+x/y}e^{-x}\ud x~D^2 \theta^{\delta}
\ud y^{-\delta}\right)\nonumber\\
&=\exp\Big(-\pi\lambda D^2 \underbrace{\theta^{\delta}\int_0^\theta \frac{\delta}{y^{\delta}}\int_{\beta}^{\infty}\frac{1}{x+y}e^{-x}\ud x\ud y}_{G(\theta)}\Big),
\label{FexLT}
\end{align}
where (a) follows from $y=\theta(D/v)^{\alpha}$. Now using (\ref{FexLT}), we can evaluate the approximation in (\ref{P1fp}) as
\begin{align}
P_1(\theta)&\approx \mathbb{E}\left[\exp\left(-\pi\lambda G(\theta) D^2 \right)\right]=\frac{1}{1+G(\theta)}\label{di_tci}.
\end{align}
The function $G(\theta)$ in (\ref{FexLT}) can be written as
\begin{equation}\label{Gaexp}
G(\theta)=\theta^{\delta}\int_0^\theta \frac{\delta}{y^{\delta}}~ e^{y}E_1\left(\beta+y\right)\ud y.
\end{equation}
From (\ref{di_tci}) and (\ref{tciDi}), the $p_{\rm s}(N)$ can be expressed as in (\ref{DiUB}).
\section{Proof of Theorem 6}
\label{sec:Theo6}
$\bar{I}\left(t\right)$ in (\ref{As_avin}) has the form of a standard Poisson additive shot noise. From \cite[Prop. 2.2.4]{Bacc_book1}, the Laplace of $\bar{I}(t)$ is
\begin{equation}\label{ItLT}
\mathcal{L}(\xi)=\exp\left(-\pi\lambda \mathbb{E}\left[\iint_D^{\infty} \left(1-e^{-\xi\abs{g}^2\bar{\eta}v^{-\alpha}}\right)\ud v^2\ud s \right]\right).
\end{equation}
The CCDF of the typical user packet transmission time $T$ in (\ref{pkt_tias}) can be derived by following steps similar to Appendix \ref{sec:Theo1}. Due to space limitations, we omit the full derivations. Only the key differences from Appendix \ref{sec:Theo1} are highlighted. From (\ref{ccf_Hu}), the CCDF can be bounded as
\begin{equation}\label{as_ccdf}
\mathbb{P}\left(T>t\right)\leq 1-\frac{1}{{}_2F_{1} \left(\left[1,-\delta\right]; 1-\delta; -\theta_t \mathbb{E}\left[\bar{\eta}\right]\right)}.
\end{equation}
The RV $\bar{\eta}$ in (\ref{as_ccdf}) has the following form
\begin{equation}\label{Eet_as}
\bar{\eta}(t)=\frac{1}{t}\int_0^t 1(S\leq \tau\leq S+\bar{T})\ud \tau.
\end{equation}
An expression for the $\mathbb{E}[\cdot]$ of $\bar{\eta}(t)$ is given by
\begin{equation}\label{Eet_asI}
\mathbb{E}\left[\bar{\eta}(t)\right]=\mathbb{E}_{\bar{T}} \int_{-\infty}^{\infty} \Big [\frac{1}{t}\int_0^t 1(s\leq \tau\leq s+\bar{T})\ud \tau \Big ] \ud s.
\end{equation}
From \cite[Appendix 2]{JourVer}, we obtain for $0<\varepsilon \leq 1$
\begin{align}\label{Eet_tde}
\mathbb{E}\left[\bar \eta(t)^{\varepsilon}\right]&=\frac{1}{N t^{\varepsilon}}\Big[\Big(t\int_0^t\bar{t}^{\varepsilon}+
\frac{1-\varepsilon}{1+\varepsilon}\int_0^t\bar{t}^{1+\varepsilon}+t^{\varepsilon} \int_t^N\bar{t}\nonumber\\
&~~~+\frac{1-\varepsilon}{1+\varepsilon}~t^{1+\varepsilon}\int_t^N\Big)\ud F(\bar{t})\Big],
\end{align}
where $F(\bar{t})$ is the CDF of $\bar T$. For $\varepsilon=1$ in (\ref{Eet_tde}), we get
\begin{equation}\label{Eet_t}
\mathbb{E}\left[\bar \eta(t)\right]=\mathbb{E}[\bar T]/N.
\end{equation}
Plugging (\ref{Eet_t}) back in (\ref{as_ccdf}) completes the proof. For the calculation in (\ref{Eet_t}), we use the same distribution of $\bar{T}$ as in (\ref{Tbacd}) of Appendix \ref{sec:Theo1}.
\bibliography{References_SRA}
\bibliographystyle{IEEEtran}
\begin{IEEEbiography}[{\includegraphics[width=1in,height
=1.25in,keepaspectratio]{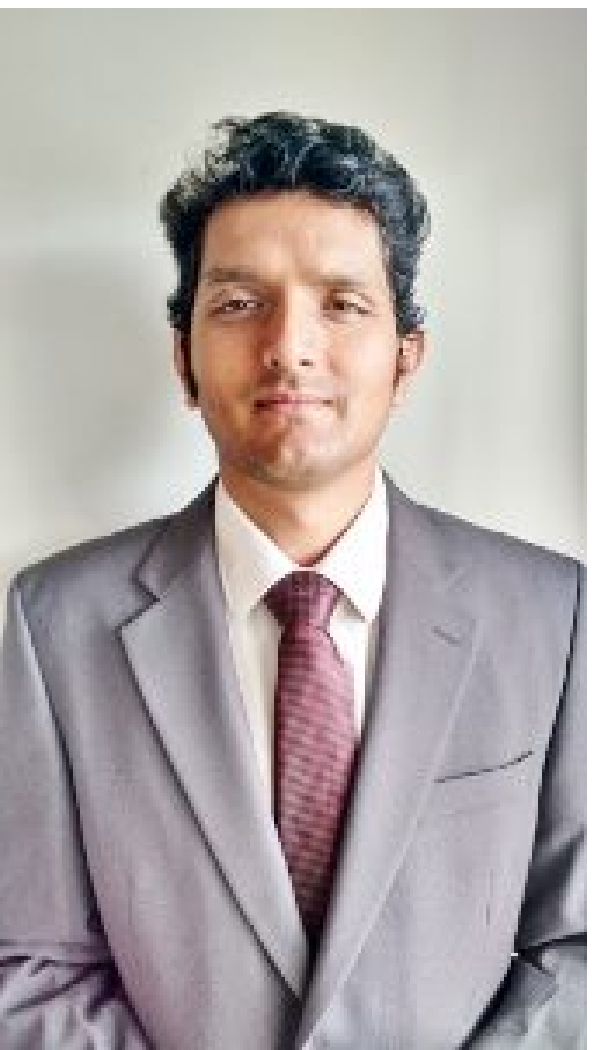}}]{Amogh Rajanna}
(M' 16) received the BS degree in electronics and communication engineering from University of Mysore, India in 2007, MS and PhD degrees in electrical engineering from University of Minnesota, Twin Cities, USA in 2011 and 2015 respectively.
He was a post-doctoral research associate at the Wireless Institute, University of Notre Dame, USA from 2015-2016. He worked as a Senior Research Associate at the Institute for Probability, Analysis and Dynamics, School of Mathematics, University of Bristol, UK from 2017-2019. Currently, he is a Visiting Researcher at the Communications Architectures and Research Section (332), Jet Propulsion Laboratory (JPL), NASA in Pasadena, CA, USA.

In 2018, he was a recipient of the EPSRC IAA (Research Council, UK) Early Career Researcher Kickstarter award which was directed at developing partnership with JPL, NASA. In 2019, he was co-awarded the EPSRC IAA Knowledge Transfer Secondment award for a project \emph{Energy-efficient Adaptive Communications Technology for the Proximity Links of NASA Space Networks} at JPL, NASA. His research interests include communication theory, stochastic geometry, information and coding theory, Markov processes applied to problems in wireless communication networks.
\end{IEEEbiography}
\begin{IEEEbiography}[{\includegraphics[width=1in,height
=1.25in,keepaspectratio]{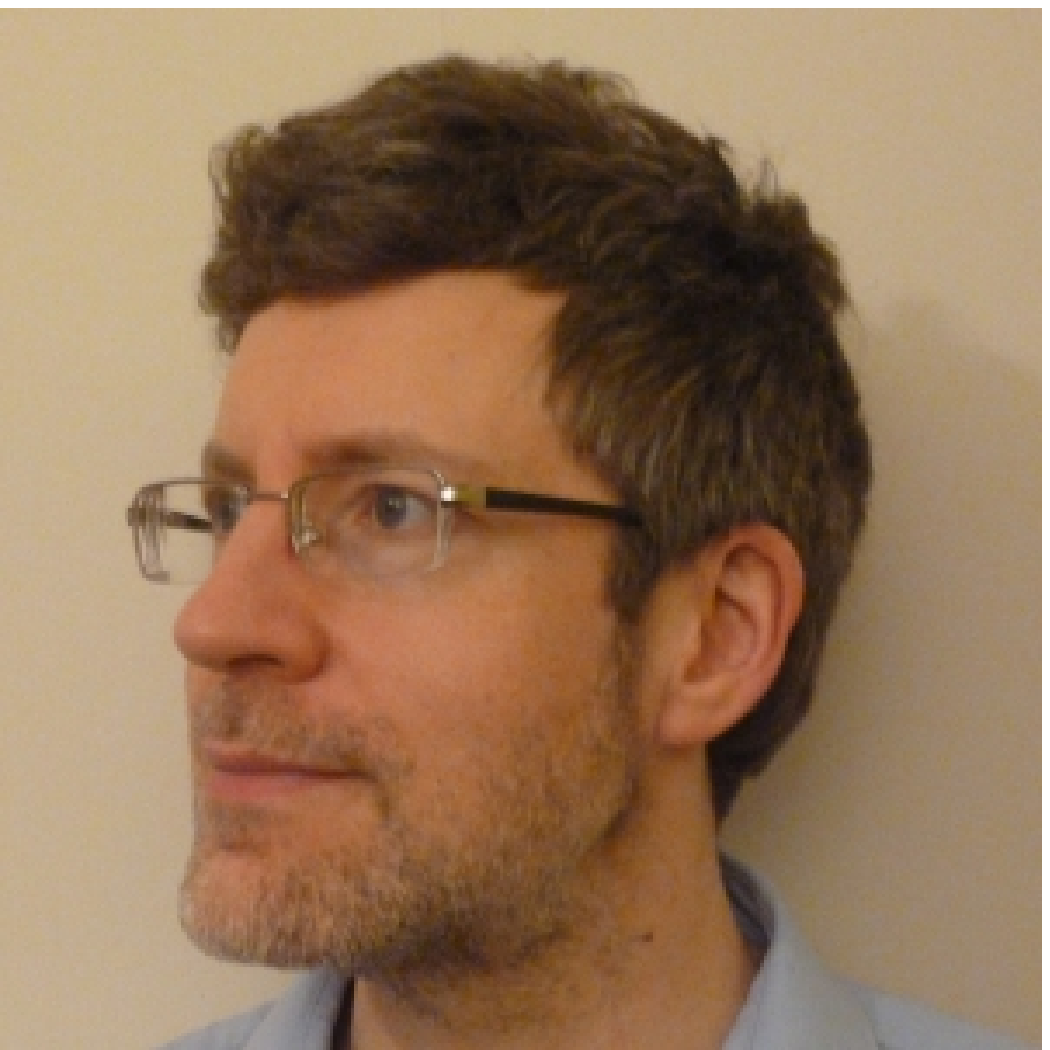}}]{Carl P. Dettmann}
received the BSc (Hons.) and the PhD degrees in physics from the University of Melbourne, Australia, in 1991 and 1995, respectively. Following research positions at New South Wales, Northwestern, Copenhagen, and Rockefeller Universities, he moved to the University of Bristol, Bristol, United Kingdom, where he is now a professor in the School of Mathematics and deputy director of the Institute of Probability, Analysis, and Dynamics. He has published more than 125 international journal and conference papers in complex and communications networks, dynamical systems, and statistical physics. He is a Fellow of the Institute of Physics and serves on its fellowship panel. He has delivered numerous presentations at international conferences, including a plenary lecture at Dynamics Days Europe and a tutorial at the International Symposium on Wireless Communication Systems.
\end{IEEEbiography}
\end{document}